\documentclass{aastex631}
\submitjournal{The Astrophysical Journal}
\received{2023 September 9}
\revised{2024 January 9}
\accepted{2024 February 4}

\definecolor{ForestGreen}{rgb}{0,.4,.1}
\definecolor{Red}{rgb}{1,0,0}
\definecolor{Purple}{rgb}{1,0,1}
\definecolor{Orange}{rgb}{1,.5,0}
\definecolor{Teal}{rgb}{0,.5,.5}

\newcommand{\nf}{\ensuremath{_{\mbox{\scriptsize nf}}}}
\newcommand{\sfeet}{\ensuremath{s_{\mbox{\scriptsize feet}}}}
\newcommand{\snf}{\ensuremath{s_{\mbox{\scriptsize nf}}}}
\newcommand{\davg}{\ensuremath{d_{\mbox{\scriptsize avg}}}}
\newcommand{\dmax}{\ensuremath{d_{\mbox{\scriptsize max}}}}
\newcommand{\tavg}{\ensuremath{\tau_{\mbox{\scriptsize avg}}}}
\newcommand{\tmax}{\ensuremath{\tau_{\mbox{\scriptsize max}}}}
\newcommand{\Iavg}{\ensuremath{I_{\mbox{\scriptsize avg}}}}
\newcommand{\Imed}{\ensuremath{I_{\mbox{\scriptsize med}}}}
\newcommand{\cc}{\mbox{cm}\ensuremath{^{-3}}}
\newcommand{\revA}[1]{{#1}}
\shorttitle{Modeling of Condensations in Coronal Loops Produced by Impulsive Heating }
\shortauthors{Kucera, Klimchuk, and Luna}


\begin{document}
\title{Modeling of Condensations in Coronal Loops Produced by Impulsive Heating with Variable Frequencies and Locations}
\author[0000-0001-9632-447X]{Therese A.\ Kucera}
\affiliation{Heliophysics Science Division, NASA Goddard Space Flight Center, Greenbelt, MD 20771, USA}
\author[0000-0003-2255-0305]{James A.\ Klimchuk}
\affiliation{Heliophysics Science Division, NASA Goddard Space Flight Center, Greenbelt, MD 20771, USA}
\author[0000-0002-3841-313X]{Manuel Luna} 
\affiliation{Departament de F\'{\i}sica, Universitat de les Illes Balears, E-07122, Palma de Mallorca, Spain}
\affiliation{Institute of Applied Computing \& Community Code (IAC$^3$), UIB, Spain}

\begin{abstract}
We present the results of models of impulsively heated coronal loops using the 1-D hydrodynamic Adaptively Refined Godunov Solver (ARGOS) code. The impulsive heating events (which we refer to as ``nanoflares'') are modeled by discrete pulses of energy along the loop. We explore the occurrence of cold condensations due to the effective equivalent of thermal non-equilibrium (TNE) in loops with steady heating, and examine its dependence on nanoflare timing and intensity and also nanoflare location along the loop, including randomized distributions of nanoflares. We find that randomizing nanoflare distributions, both in time/intensity and location,  diminishes the likelihood of condensations as compared to distributions with regularly occurring nanoflares with the same average properties. The usual criteria that condensations are favored for heating near loop footpoints and with high cadences are more strict for randomized (as opposed to regular) nanoflare distributions, and for randomized distributions the condensations stay in the loop for a shorter amount of time. \revA{That said, condensations can sometimes occur in cases where the average values of parameters (frequency or location) are beyond the critical limits above which condensations do not occur for corresponding steady, non-randomized values of those parameters.} These properties of condensations occurring due to randomized heating can be used in the future to investigate diagnostics of coronal heating mechanisms.
 \end{abstract}
 
 \section{Introduction}
 \label{s:intro}

Although in general the solar corona is well known to be hot ($\ga 10^6$~K) there also exists in the corona much cooler (~$10^4$~K) plasma observed in such common phenomena as solar prominences and coronal rain.  Thermal non-equilibrium \citep[see recent review by][]{antolin_22} is thought to be a key process involved in the formation of such cool plasma in the corona, and understanding it and how it operates under realistic scenarios can help us to better understand and diagnose coronal heating processes.

 Thermal non-equilibrium (TNE) occurs because the lower legs and upper part of a coronal loop are unable to achieve a local energy balance. This is the situation when steady coronal heating is concentrated at low altitudes. The legs require large density for radiation to balance the strong heating, which leads to the evaporation of chromospheric plasma into the loop.  The up-flowing plasma does not remain confined to the lower legs, however, and the resulting density increase in the upper portion of the loop leads to radiative losses that exceed the heating there. A thermal runaway ensues, and a cold condensation is formed. Note that the thermal runaway is closely related to but different from a thermal instability \citep{klimchuk_19}\revA{, although see \citep{waters_23} for a different perspective}. 
  
\revA{Thermal non-equilibrium strictly refers to loops with steady heating, but impulsive heating events can also give rise to the same basic condensation process.} The difference is that material drains from the loop between heating events. Density oscillates rather than increasing monotonically, as it does with steady heating. If the heating events are frequent enough, the peak density increases with each successive event, and eventually a thermal runaway occurs. If the events have a regular cadence, there is a minimum frequency that must be achieved for this to happen \citep[e.g.,][]{karpen_08,susino_10}. \revA{We here further investigate the conditions under which impulsive heating can lead to condensation formation.}

Location of heating also  influences the density growth required for \revA{thermal runaway}. Heating location determines the strength (mass flux) of the evaporation. Heating closer to the feet of the loop produces a steeper temperature gradient, greater conduction flux, and stronger evaporation. Just as with steady heating \citep{serio_81}, there is a requirement on how low in the loop legs the heating must occur \citep[e.g.,][]{testa_05,susino_10}, and high levels of background heating throughout the loop inhibits condensations \citep[see, e.g.,][]{johnston_19b,klimchuk_luna_19}. Furthermore, \revA{asymmetries in the heating and/or loop geometry can be very important. While expansion of the cross-sectional area with height promotes TNE, asymmetries in the area or the heating tend to inhibit TNE \citep{klimchuk_luna_19}. However, different asymmetries can sometimes counter balance each other, allowing condensations to form when they might not otherwise  \citep{froment_18,pelouze_22}.}
 
Here we present a series of one-dimensional loop simulations exploring the effects of variations in the  timing and location of impulsive heating events on loop condensations. In Section~\ref{s:numerical_simulations} we describe the code used and the general properties of the simulations and the loops. Section~\ref{s:sim_results} describes the results of the individual simulations, including ones of symmetric heating and heating alternating between legs for constant and randomly varying time and intensity (Section~\ref{s:var_delay}) and constant and randomized locations (Section~\ref{s:var_location}). Finally, we investigate cases in which the heating is randomized for both delay time and for location (Section~\ref{s:rand_delay_rand_loc}). Further discussion and comparison with other work are presented in Section~\ref{s:discussion}, and the results are summarized in Section~\ref{s:summary}.
 
 \section{Numerical Simulations}
 \label{s:numerical_simulations}

To study the effects of distributions \revA{of} impulsive heating events on condensations we modeled a series of 1D hydrodynamic loops using the Adaptively Refined Godunov Solver (ARGOS) \citep{antiochos_99}. ARGOS has an adaptively refined mesh, and we adopt a 4.62 km minimum cell size. 

 \subsection{Model Loop Geometry}
 \label{s:loop_geom}
 \begin{figure}
\includegraphics[height=8cm]{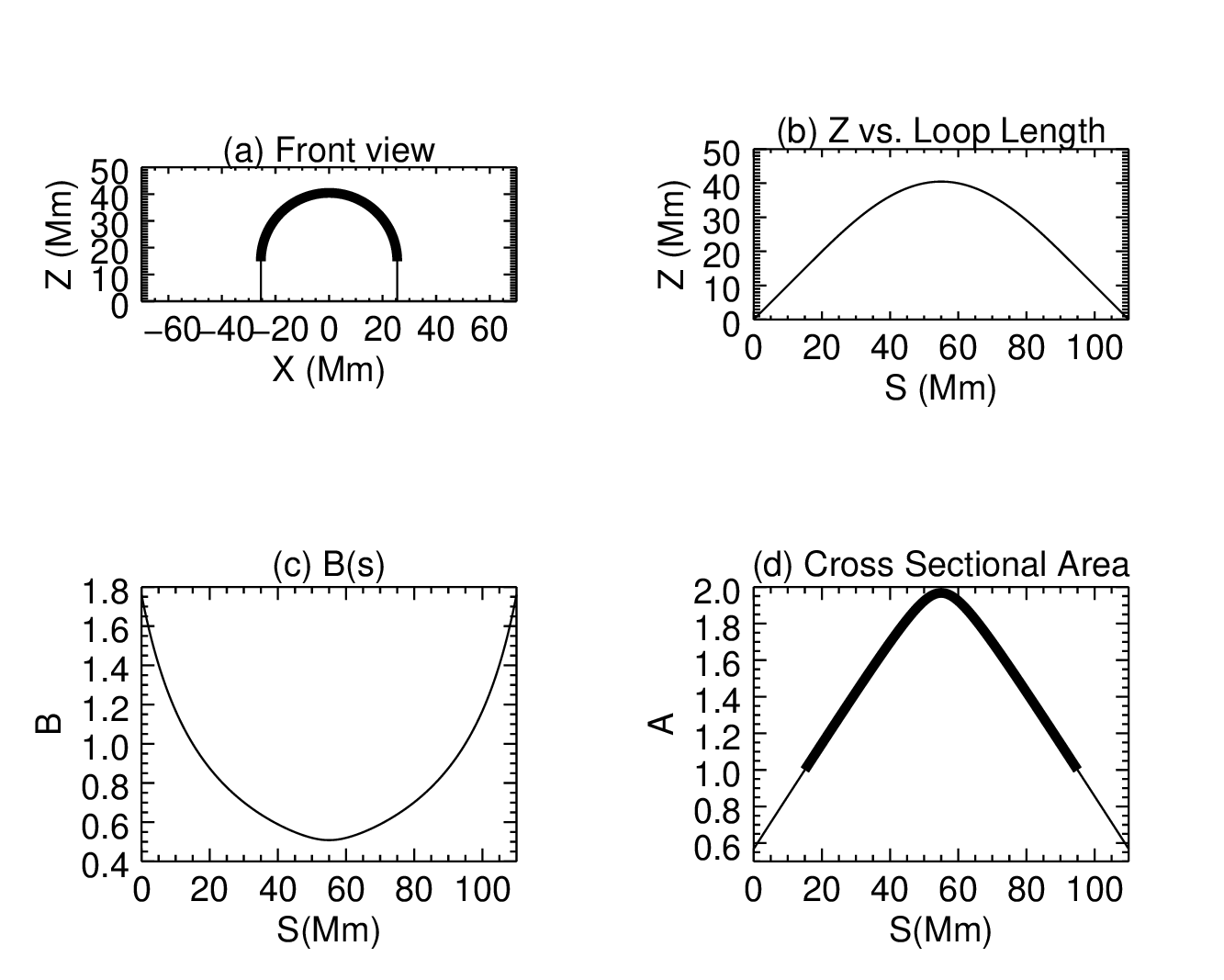}
\caption{Parameters for the main, $L_0=80$~Mm,  loop used in our modeling. (a) The view of the loop in the x-z plain; (b) altitude, $z$, as a function of the distance along the loop, $s$; (c) the normalized magnetic field strength as a function of $s$; and (d) the normalized cross sectional area of the loop as a function of $s$. The bolded portion of the lines show the region corresponding to $L_0$, which forms a semicircle.}
\label{f:LoopA}
\end{figure}

Our model loop is a semi-circle with a vertical section at each end (see Figure~\ref{f:LoopA}). The semicircular portion has length $L_0$, and the vertical feet at each end have length \sfeet. The footpoint regions (or ``feet'') are initially chromospheric in character, but the top of the chromosphere moves dynamically through the feet in response to the nanoflare heating over the course of the simulations. The loop chromosphere continues on for 40~Mm below the loop feet and is not shown in the figures in this paper. 

In most of our runs the initial semi-circular portion of the loop is $L_0=80$~Mm, with an additional $\sfeet=15$~Mm of vertical flux tube on on either side.  In addition, for a few runs discussed in Section~\ref{s:const_delay} we use a longer loop with $L_0=130$~Mm.

For the model loop we assume the cross-sectional area is inversely proportional to the magnetic field so that it expands with altitude.  
The entire loop, including the feet, expands as
$$A= C_n(1 - 0.1\sqrt{(1+(s-L_0/2)^2/7.5^2)}),$$
where $C_n$ is a normalization constant, equal to 2.19 for $L_0=80$~Mm and 2.91 for $L_0=130$~Mm, set so that the area equals unity at the base of the semicircular portion of the loop and expands by a factor of two by the apex. 

\begin{figure}
 \includegraphics[height=6cm]{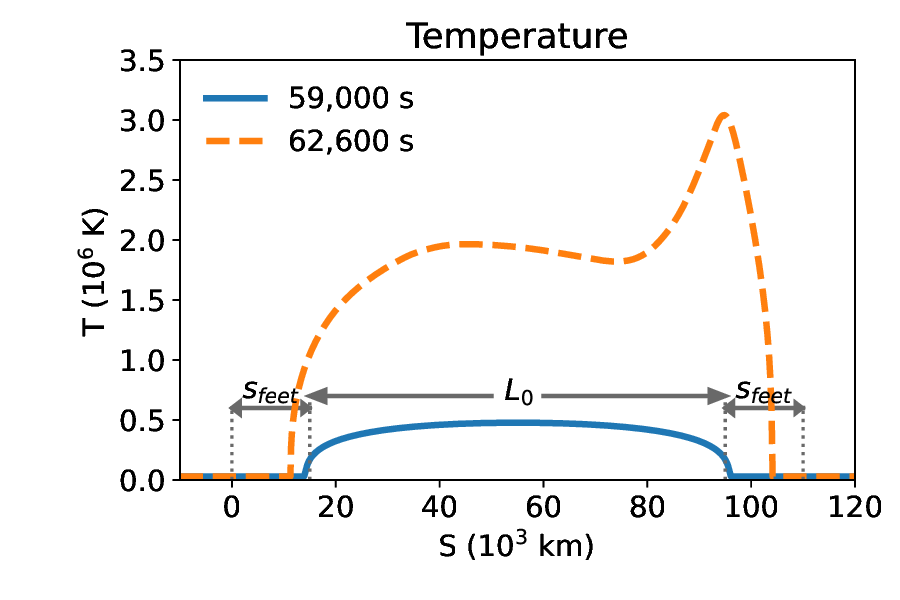}
\caption{Temperature profiles for a loop at its initial equilibrium (blue solid) and during a nanoflare 15 Mm from the base of the right-hand foot, i.e., at the initial top of the chromosphere (orange dotted). Also shown are the length of the coronal portion of the loop at the start of the simulation, $L_0$, and regions through which the coronal plasma expands once nanoflare heating commences, \sfeet.}
\label{f:TProfileExamples}
\end{figure}

 \subsection{Loop Heating} 

 We use the radiative loss function of \citet{klimchuk_08} with the exception of a $T^3$ dependence below 0.1 MK to account approximately for optical depth effects and an abrupt decrease below $3\times10^4$~K to maintain a nearly isothermal chromosphere at that temperature, as done in \citet{klimchuk_10}.

\begin{equation}
\Lambda(T) = \cases{5.49\times10^{-16} T^{-1} 			& $T > 10^{6.9}$\cr
				 3.46\times10^{-25} T^{1/3}		& $10^{6.55}<T\le10^{6.9}$ \cr	
				 3.53\times10^{-13}  T^{-3/2}		& $10^{6.18}<T\le10^{6.55}$ \cr
				 1.90\times10^{-22} 				& $10^{5.67}<T\le10^{6.18}$\cr
				 8.87\times10^{-17} T^{-1}			& $10^5<T\le10^{5.67}$\cr  
				8.87\times10^{-37} T^3			& $3\times10^4<T\le10^5$\cr 				 
				4.79\times10^{-26}(T-2.95\times10^4)& $2.95\times10^4<T\le3\times10^4$\cr
				 0							& $T\le2.95\times10^4$ }
\end{equation}

We start with  steady uniform heating of $2.4\times10^{-6}$ erg~cm$^{-3}$~s$^{-1}$ maintained over the length of the loop including the feet.  This value was chosen based on equilibrium loop scaling laws to achieve a desired coronal temperature. We then allow  the atmosphere to relax for 60,000 s and establish a proper equilibrium with peak temperature of about $5\times10^5$~K, as shown by the blue curve in Figure~\ref{f:TProfileExamples}.
The steady uniform heating is maintained as the impulsive heating events are later imposed. We also include an additional temperature and density dependent heating in the chromosphere to maintain the desired chromospheric temperature in the presence of expansion cooling \citep[see discussion in][]{sow_mondal_22}.

After $t=60,000$~s we introduce heating in short confined pulses we refer to as ``nanoflares'' although there is nothing specifying the physical origin of the heating events. They could be magnetic reconnection based or wave based. The heating is expected to be impulsive in both cases \citep{klimchuk_06}. As mentioned above, when additional nanoflare heating is added to the loop, the top of the chromosphere moves downward through the 15~Mm loop ``feet''  so that the top of the chromosphere is usually below the initial equilibrium location. 

 The nanoflare heating events are represented by triangular pulses in time and location in the loop. Each nanoflare is characterized by a duration and width  specifying the base of the triangles in the temporal and spatial dimension. For all our runs the nanoflare duration=100~s and width=5~Mm. The peak intensity, $I\nf$, time of peak heating, $t\nf$,  and location of peak heating, $s\nf$, were varied as described below.
 
 The period \revA{of time with} nanoflare heating was usually extended to about t=200,000~s (140,000~s after the end of the equilibrium run), but, because the first nanoflares of each run did not always start at exactly the same time, the standard time extent used in our analysis of the results is 135,000~s starting 300~s after the peak of the first nanoflare. We had a few runs that were less than 200,000~s either because they were producing \revA{condensations} at a very regular cadence, making a longer run unneeded, or because the run crashed. 
 
\subsection{Model Run Input and Output Parameters}
\label{s:param}
  
Input parameters varied between different models runs include $\tau$, the delay time between successive nanoflares; $ I\nf$, the nanoflare intensity; and 
$s\nf$, the location of the center of the nanoflare measured from the bottom of the left loop foot. 

 The output data from the runs include temperature, electron density, and velocity at all locations along the loop.
 The state of the model was output with a 50~s cadence, and for analysis the parameters were interpolated onto a regular spatial grid with 190~km cells. In order to characterize our runs we have derived other parameters from these values.

\begin{figure}
 \includegraphics[height=5.5cm]{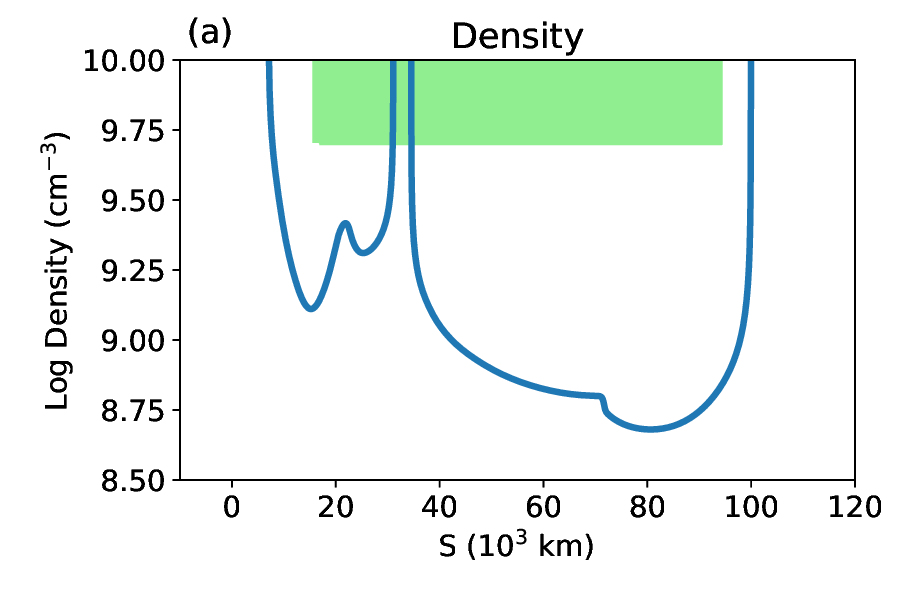}
 \includegraphics[height=5.5cm]{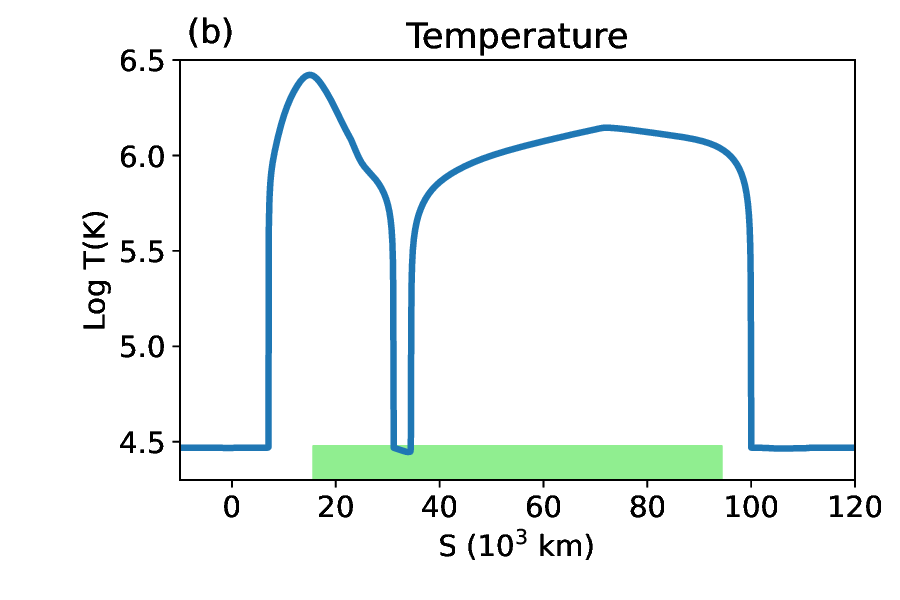}
\caption{Loop (a) density and (b) temperature showing a condensation in a model loop with $L_0=80$~Mm. The green areas show the parameter ranges used to determine the presence of condensed material}
 \label{f:cond_ranges}
 \end{figure}

\begin{figure}
  \includegraphics[height=8cm]{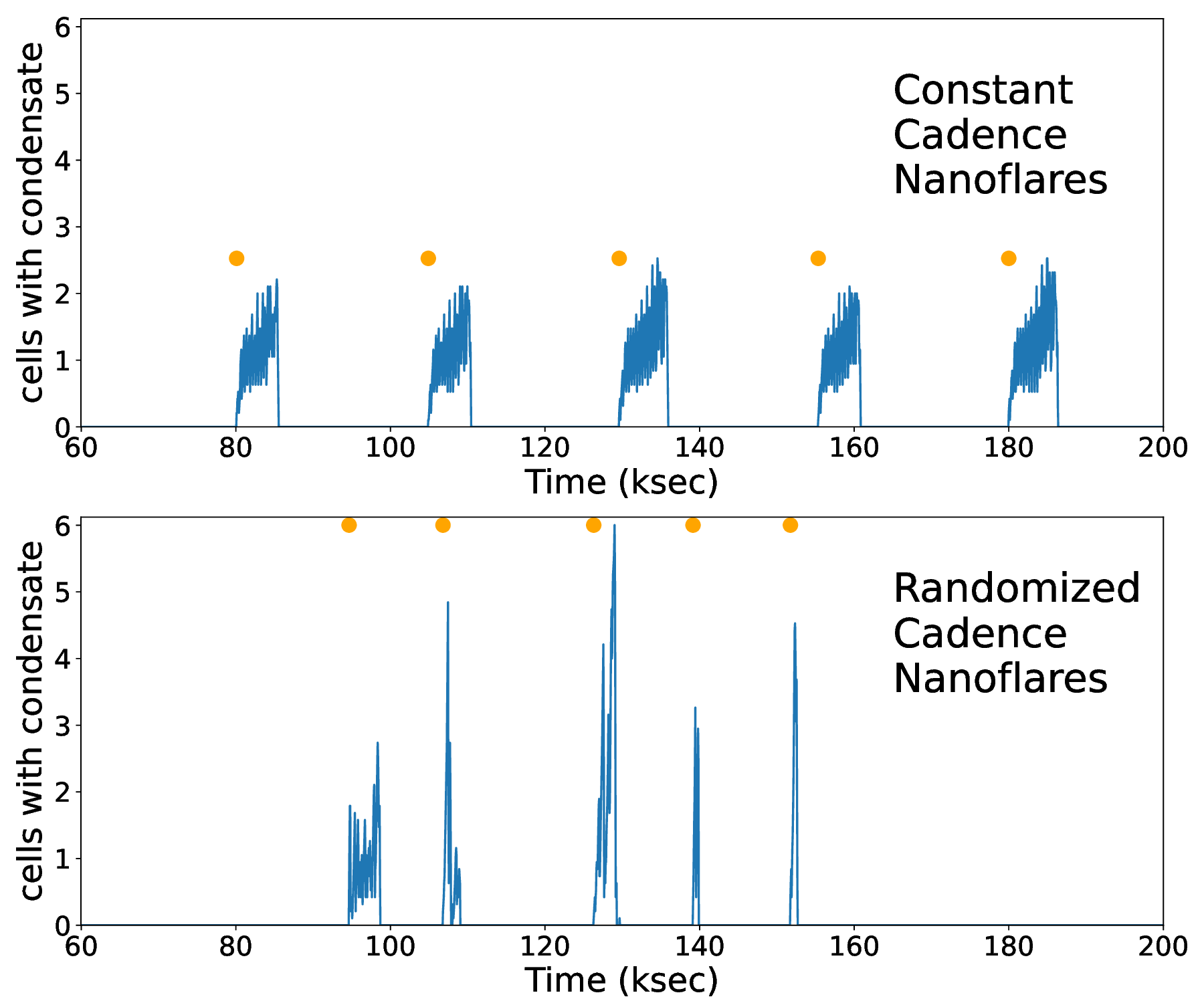}
  \caption{Examples of plots showing the number of loop pixels containing condensate as a function of time for a run with nanoflares with a regular cadence (top) and a randomized cadence (bottom). Orange dots show the start of each condensation event. The width of the feature corresponds to the duration and the height to the size of the condensation. High frequency fluctuations within  the condensations are an artifact of the automatic detection method. Monotonic trends are real (e.g., condensations growing in time).}
 \label{f:cond_count}
 \end{figure}

 A key output parameter we consider is the number of condensations that are formed per unit time. For the model loops with $L_0=80$~Mm condensations are automatically counted by detecting all pixels along the loop between \revA{$s = 15.7$ and 94.3~Mm}  with $n_e>10^{9.7}$~\cc\ and $T<30,000$~K,
  as shown in Figure~\ref{f:cond_ranges}. For model loops with $L_0=130$~Mm we used $s$ between 15 and 145~Mm. The range of $s$ is selected to be as large as possible without including area through which the top of the chromosphere moves. It does mean that some cool features very close to the end of the loops are not counted. These features are usually small and fall directly and rapidly down into the chromosphere. This technique results in a time series showing the number of pixels containing condensate. This value fluctuates as the density and temperature move outside of our cut-offs in $n_e$ and $T$ so we smooth the result with a 2,000~s boxcar function and the resulting features with one or more pixels showing cool, dense plasma are counted as a single condensation, allowing us to automatically characterize the condensations. Examples of the number of pixels with condensate as a function of time are shown in Figure~\ref{f:cond_count}. Because the condensations are episodic with a relatively small number of condensations in each model run ($\le10$) the number is not exact, but precise condensation rates are not our goal. There are some uncommon cases in which there are two condensations in a loop at the same or nearly the same time, and these condensations are counted as a single condensation. In addition, we discuss the average duration of condensations, also derived from the time series of the number of pixels containing condensate.
 
Another output parameter is the average distance of the nanoflares from the dynamic top of the chromosphere at run time, \davg. This is distinct from \revA{distance} of the nanoflares from the bottom of the loop feet, which is an input parameter.  For some of the runs with randomized locations (see Section~\ref{s:rand_location}) there were nanoflares that occurred below the top of the chromosphere at the time of the nanoflare. A nanoflare occurring entirely below the chromosphere does not result in chromospheric evaporation, so we excluded nanoflares centered more than 1 Mm below the top of the chromosphere from our calculations of \davg\ and also \tavg, the average time between events.
Because the nanoflares have a triangular spatial profile with a 5 Mm base this excludes only nanoflares mostly or entirely submerged in the chromosphere. It keeps nanoflares where at least 18\% of the energy is deposited above the chromosphere. For simulations with nanoflares submerged in the chromosphere, this calculation results in longer \davg\ and \tavg\ values.  

\revA{Lists of the model runs with key input and output parameters are presented in the tables in Appendix~\ref{s:tables}.}

 \section{Simulation Results}
 \label{s:sim_results}
 \subsection{Variation in Delay Between Nanoflares}
 \label{s:var_delay}
  
 \subsubsection{Nanoflare Distributions with Constant Delay Time and Intensity}
 \label{s:const_delay}
 We started with a study of the effects of nanoflares released at regular cadences (constant $\tau$).  We considered both cases in which nanoflares occurred simultaneously in both legs of the loop and cases in which the nanoflares alternated between the two legs. 
 
 We define an ``event'' as the input of energy into the loop at a given time, so that a single nanoflare can be an event but two nanoflares occurring at the same time are also considered a single event. For the runs described in this section the \revA{peak} energy in each event was set to 0.2~erg~s$^{-1}$\cc. Thus if there was a single nanoflare (as in the alternating or randomized delay cases) it would have a peak energy of 0.2~erg~s$^{-1}$\cc, but for runs in which two nanoflares occurred simultaneously the peak energy was 0.1~erg~s$^{-1}$\cc\ in each nanoflare so that the entire nanoflare energy over the loop would be the same in the symmetric and alternating cases.  For a single nanoflare with $I\nf=0.2$~erg~s$^{-1}~\cc$ the total additional energy integrated over the loop and the duration of the nanoflare is $2.5\times10^9$~erg~cm$^{-2}$.

 \revA{Each nanoflare triggers a flow of higher density material into the loop, which, in absence of other nanoflares, can bounce back and forth between loop legs before dissipating. As we discuss below, however, if the conditions are right such flows from multiple nanoflares can interact, leading to a buildup of density that becomes sufficiently high as to result in catastrophic cooling.} 

 As discussed in Section~\ref{s:numerical_simulations}, our main model loop was $L_0=80$~Mm long, not including a 15~Mm section at each end that was initially chromospheric in character, but through which the top of the chromosphere moved once the nanoflare heating commenced. Nanoflares for these runs with varying delay times were located at $s=15$ and 95~Mm, so at the top of the chromosphere at the start of the run.  Once the nanoflaring started the height of the chromosphere moved downwards to differing extents as a function of time and the simulation characteristics. 
For symmetric, simultaneous nanoflare runs discussed here \davg\ ranged from 5.7- 8.4~Mm,  increasing with decreasing delay time as the higher frequency of heating pulses pushed down the chromosphere.  For the alternating nanoflare simulations \davg\ ranged between 5.8 - 16.2~Mm and was not monotonic with delay time, reflecting condensation rates, which will be discussed below.
    
\begin{figure}
\includegraphics[height=5.5cm]{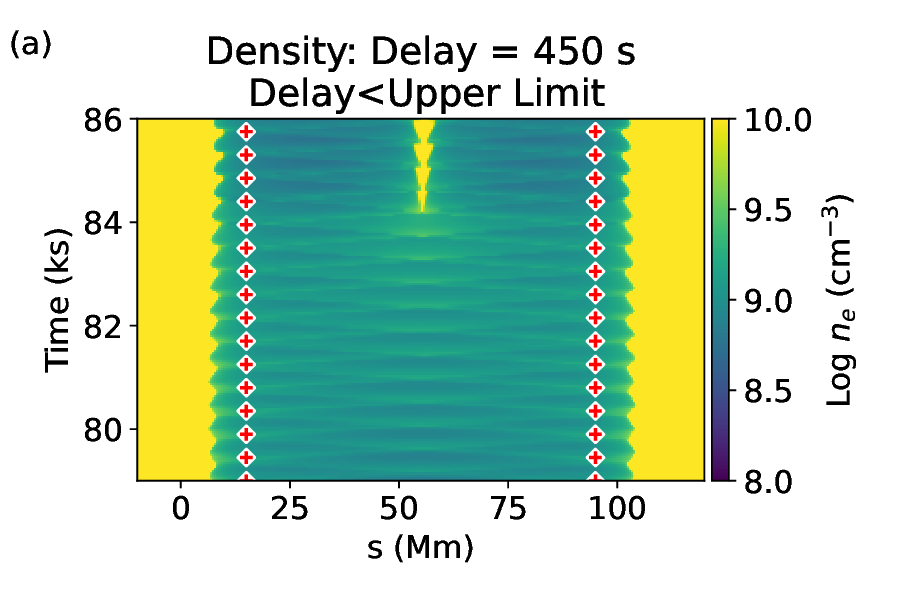}
\includegraphics[height=5.5cm]{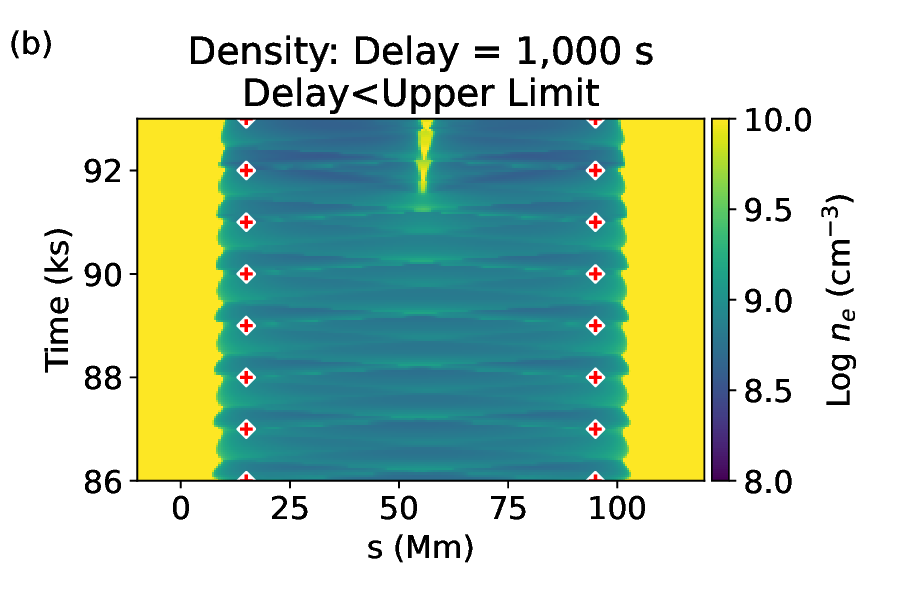}\\
\includegraphics[height=5.5cm]{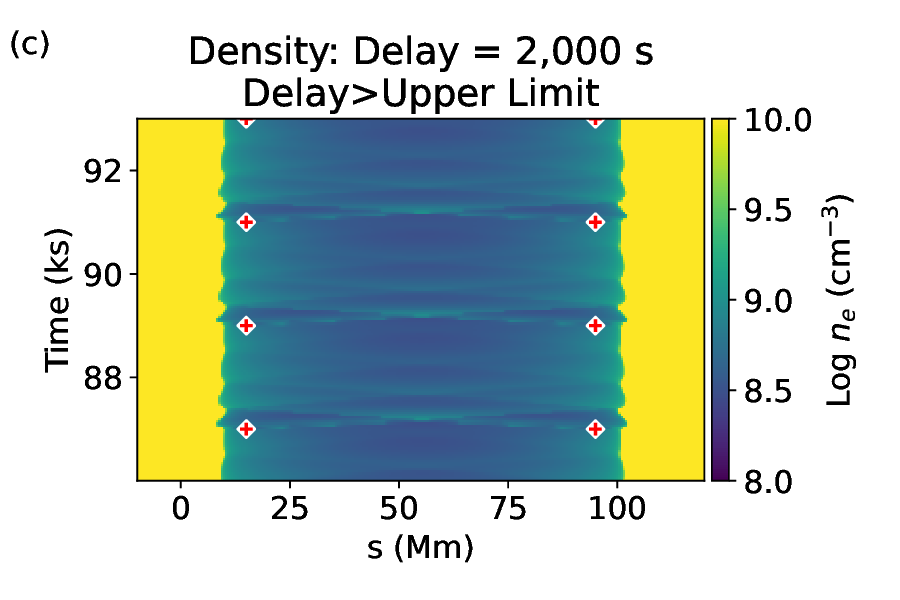}
\caption{Density as a function of location along the loop and time for three model runs with nanoflares occurring simultaneously and symmetrically in the two legs with constant delay times and locations, but variations in the delay time, $\tau$, (450, 1,000, and 2,000~s respectively) between model runs. The simulations shown in (a) and (b) have $\tau$ below the critical delay time, \tmax\ , while in (c) the delay time, 2,000~s is above \tmax. The times and locations of the nanoflares are shown by the red plus signs (+). }
\label{f:symetric_density}
 \end{figure}

 For the simulations with symmetric, simultaneous nanoflares the condensations form at the apex of the loop and are relatively stationary (see Figure~\ref{f:symetric_density}(a) and (b)) until small numerical asymmetries cause them to move to one side and fall out of the loop. Then a new cycle begins. In the figure we only show the condensation phase. 
 
\begin{figure}
\includegraphics[height=5.5cm]{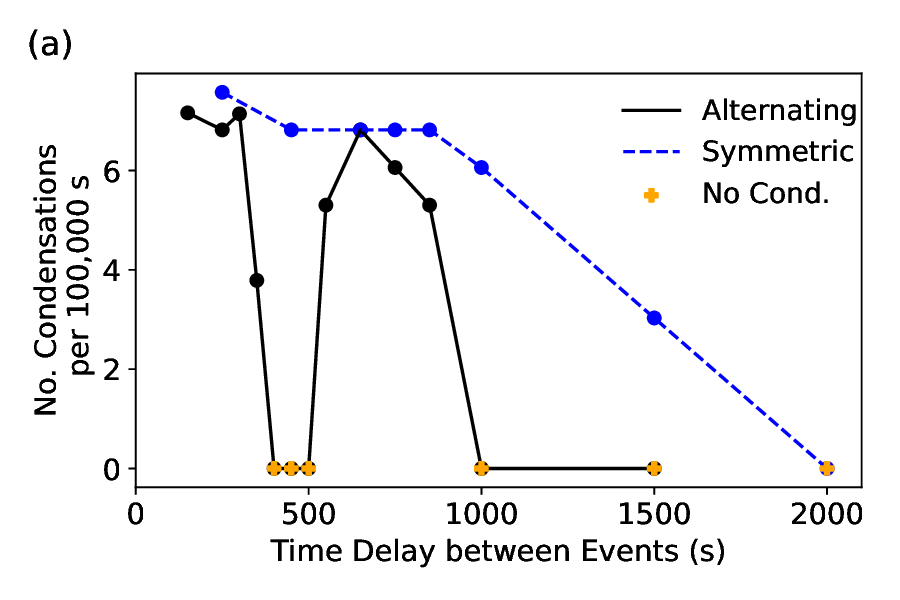}
\includegraphics[height=5.5cm]{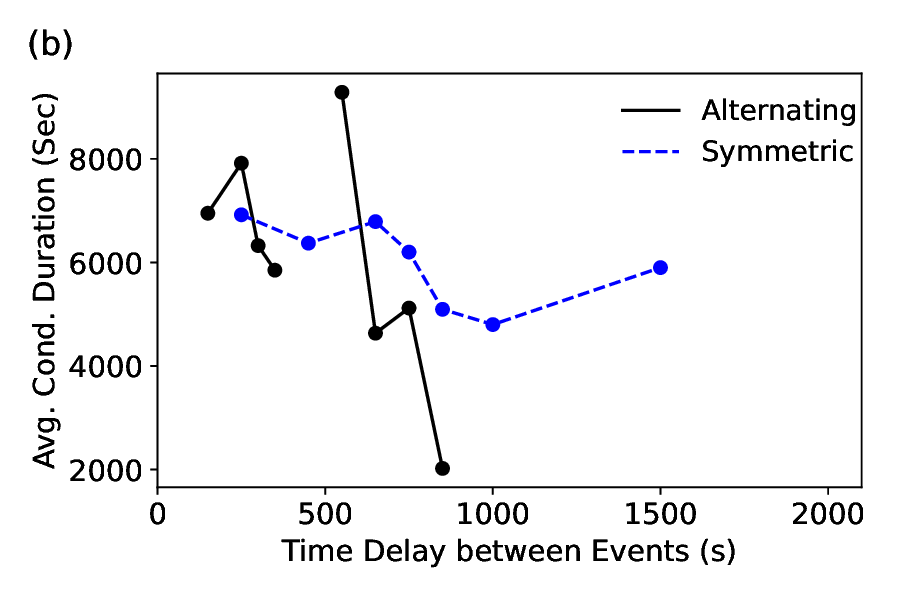}
\caption{Condensation characteristics for runs with regularly occurring nanoflares occurring simultaneously in both legs (blue dashed line) or alternating between legs (black solid line). (a) The number of condensations per 100,000~s, (b) average duration of condensations over the run. These runs were performed on a loop with $L_0=80$~Mm. \revA{Data for this figure are in Table~\ref{t:sym_alt_rand_ncond}.} }
\label{f:sym_alt_ncond}
\end{figure}

\revA{Figure~\ref{f:sym_alt_ncond} (with data in Table~\ref{t:sym_alt_rand_ncond}) shows the dependence of condensation formation frequency and duration on delay time.} 

We find that for the case of symmetrically placed, simultaneous nanoflares the results were as expected from previous work, with high incidences of condensation for cases with short delays between nanoflares, approximating constant heating.  This is shown by the dashed line in Figure~\ref{f:sym_alt_ncond}a.
For runs with $\tau\ga 1,000$~s the frequency of condensations decreases until they no longer form for delays greater than about  $\tau\ga 2,000$~s, as shown in Figure~\ref{f:symetric_density}c). 

This effect is the result of the density rising and falling during each nanoflare cycle, with the peak density increasing with each successive event. The plasma cooling time decreases with increasing density, and eventually it may become short enough for a condensation to form. If $\tau$ is too long (i.e., the nanoflare frequency is too low), these conditions may never be met because there is so much draining between events that density never builds to substantial levels. Thus, for a given nanoflare energy and height, there is a critical delay, \tmax, above which condensations do not form.   Below \tmax, the loop locally reaches conditions where the radiative cooling time is shorter than the draining time, whereas above \tmax\ the cooling time is always longer than the draining time. Furthermore, the build-up to critical density is faster with more frequent nanoflares, so shorter  $\tau$ (higher nanoflare frequencies) are associated with higher condensation rates for $1,000 <\tau <2,000$~s. Below 1,000 s the heating is effectively steady, and there is a maximum condensation rate determined by the physics of TNE cycles. The small differences in the condensation rates below \revA{$\tau=$}1,000 s in Figure 6a are due to variable condensation lifetime associated with the random numerical effects that determine when the condensation falls from the loop apex. Also, the time averaged heating rates are different for the different delays because the nanoflare intensities are the same. \revA{As can be seen from Figure~\ref{f:cond_count}a, the cadence of condensation formation for simulations with constant nanoflare frequency and location is very regular. The shortest delays correspond to a condensation recurrence period of about 3.7 hours. This is roughly similar to the period seen by \cite{winebarger_18} for a somewhat similar loop, although they were observing incomplete condensations, not the topic of this paper.}

For the alternating nanoflares runs the condensations also form near the loop apex and then the condensations move back and forth in response to the alternating heating (Figure~\ref{f:alternating_density}) with a range that increases as the delay between nanoflares increases. 
As shown by the solid line in Figure~\ref{f:sym_alt_ncond}a, \tmax\ for simulations with nanoflares alternating between legs is about 1,000~s, shorter than for the runs with simultaneous nanoflares in both legs, for which $\tmax\approx 2,000$~s. 

Figure~\ref{f:sym_alt_ncond}b shows the durations of the condensations as a function of $\tau$. The duration of the condensation for the symmetrically nanoflaring simulations varies somewhat with delay time, but not systematically. The variation in duration for the simulations with alternating nanoflares is larger.
As mentioned above, for the simulations with alternating nanoflares the condensations also form near the loop apex and then move back and forth in response to the alternating heating (Figure~\ref{f:alternating_density}).  Because the motion is caused by the nanoflares, $\tau$ effects the time the condensation stays in the corona. For instance, for $\tau=550$~s, immediately above the resonance, the condensation is trapped in the loop for a relatively long time between the alternating nanoflares.

 \begin{figure}
\includegraphics[height=10cm]{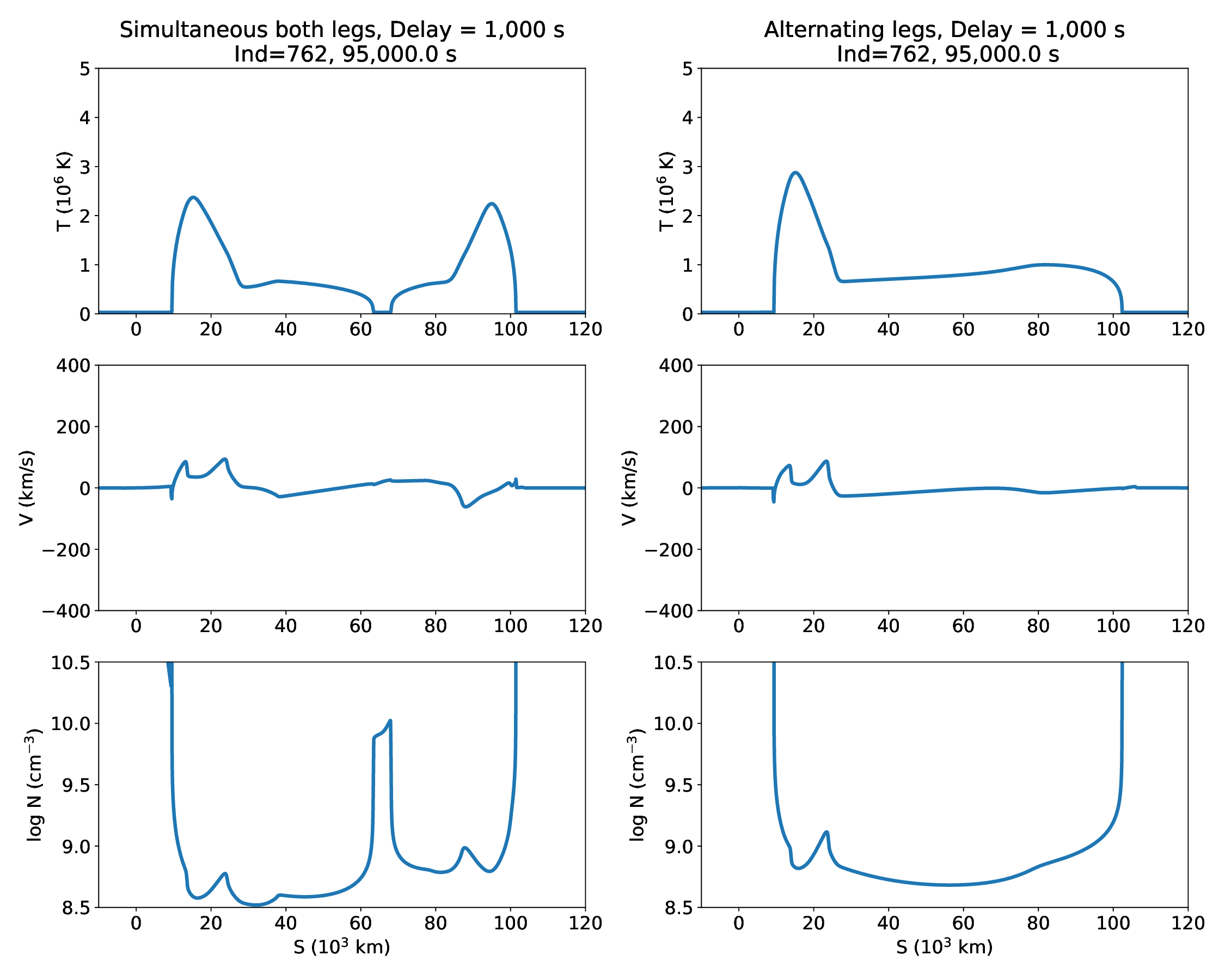}
\caption{Temperature, velocity, and density along the loop for (left) simulation with symmetric nanoflaring in both legs every 1,000 sec and (right) simulation with alternate nanoflares between legs so that there is a nanoflare in the loop every 1,000 sec. A 29 second animation showing the simulation results from 80,900 - 115,000~s is included online.}
\label{f:NF4_20Dec2021_c2_10Mar2021_c7_frame}
\end{figure}

So, based on these simulations, we find that symmetric heating enhances condensation formation compared to asymmetric heating \revA{in loops with symmetric geometry (shape and cross-sectional area)}. \revA{This is consistent with previous studies \citep{mikic_13,klimchuk_luna_19,froment_18,pelouze_22}}. For a given global nanoflare event frequency and time averaged heating rate, the build-up of apex density is faster with simultaneous nanoflares in the two legs than with alternating nanoflares, due to the colliding evaporation from the two sides. When a nanoflare occurs on one side only, material flows over the apex to the other side. Although there is a rebound, the overall increase in density is reduced.  Condensation rates are therefore faster for simultaneous nanoflares. In addition, because a higher peak density can be achieved with simultaneous nanoflares, \tmax\ is higher (i.e., the critical frequency for the formation of condensations is lower).
This is illustrated in the video accompanying Figure~\ref{f:NF4_20Dec2021_c2_10Mar2021_c7_frame} in which we show the model runs with simultaneous symmetric nanoflares and alternating nanoflares with the delay and intensity such that the same amount of energy is being release every 1,000~s. Condensations form in the symmetric, simultaneous nanoflare run, but not in the simulation with the alternating nanoflares.

It turns out that for the conditions we have considered (loop length, nanoflare energy, etc.) the critical \textit{single-leg} frequency (as opposed to the event frequency in the entire loop)  for condensation is similar for simultaneous and alternating nanoflares. This is not a general property, though, as can be seen from the analysis of a longer loop presented below.

Another difference between the simultaneous and alternating simulations is that under special circumstances – when nanoflares alternate regularly between sides at a particular frequency – a resonance occurs in which material “sloshes” back and forth, and no condensation can form. This we see in Figure~\ref{f:sym_alt_ncond}a for
$\tau$ from 400-500~s in which condensations form in the symmetric nanoflare runs but not for the alternating nanoflare simulations. This gap near $\tau\approx 450$~s is based on the time it takes the density pulse associated with each nanoflare to cross the loop. It also means the nanoflares are in sync with the motion of the top of the chromosphere so that they are uniformly occurring near the time of lowest extent of the chromosphere, resulting in a relatively large \davg\ values between 12.5-16.2~Mm. This resonance phenomenon would occur rarely if ever on the Sun, but it can help shed light on the processes at work.

The different delay regimes for alternating nanoflares are illustrated in Figure~\ref{f:alternating_density}. It shows the density profiles over about 7,000~s for four different runs of the $L_0=80$ Mm loop. We see in panel (b) showing the simulation with $\tau=450$~s between alternating nanoflares the evaporation density front associated with each nanoflare crosses the loop in such a way to reach the other side just as the next nanoflare occurs. The density does not increase as the cycle repeats. This is not the case for the runs done with nanoflares occurring at shorter (panel a) or longer delay (panel c) times. In these cases, the flows initiated at the two ends interact with each other, producing a density enhancement up in the loop. In these the density rises and falls during each nanoflare cycle, with the peak density increasing with each successive event until, eventually, the density builds up enough to result in runaway cooling and a condensation.  This a less efficient version of the density buildup that occurs when flows collide at the loop apex when there are simultaneous and symmetric nanoflares. For simulations with $\tau\ga 1,000$~s (panel d) the effects of each nanoflare have effectively disappeared by the time the next nanoflare occurs and the density never builds up sufficiently to result in a condensation, as discussed above. Although we have discussed the evolution leading up to a condensation in terms of increasing density, decreasing temperature also plays a role. Temperature oscillates, like density, but there is an overall decrease, which increases the radiative cooling.

\begin{figure}
  \includegraphics[height=5.5cm]{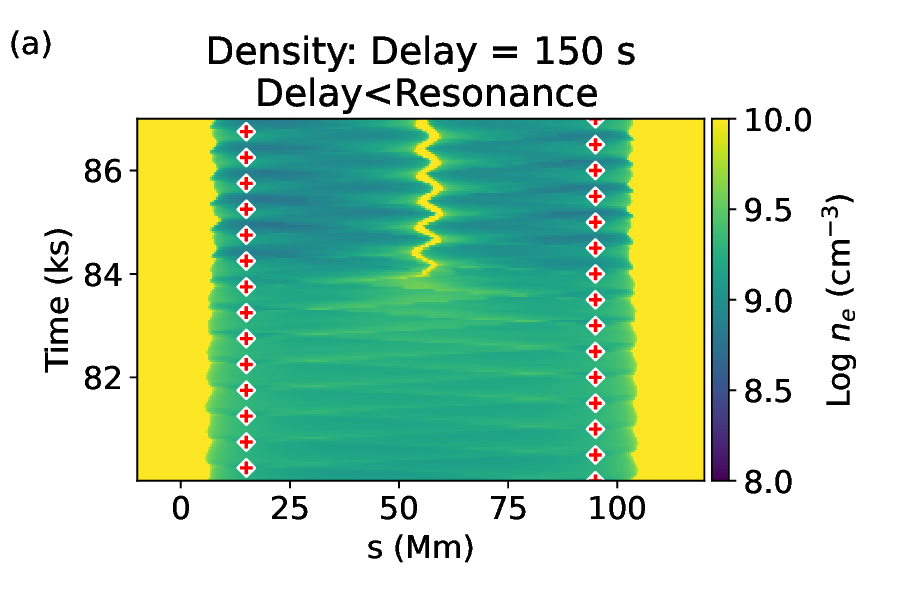}
  \includegraphics[height=5.5cm]{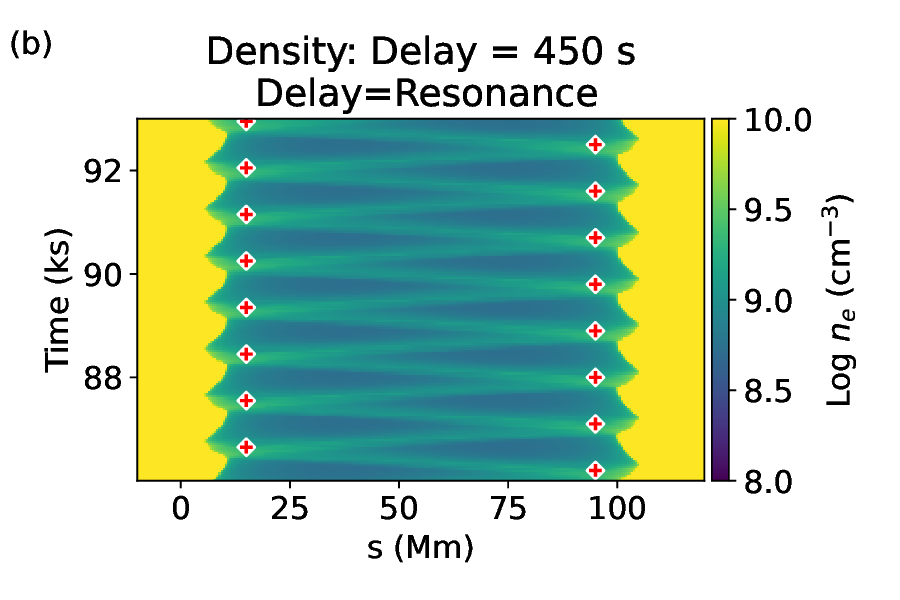}\\
  \includegraphics[height=5.5cm]{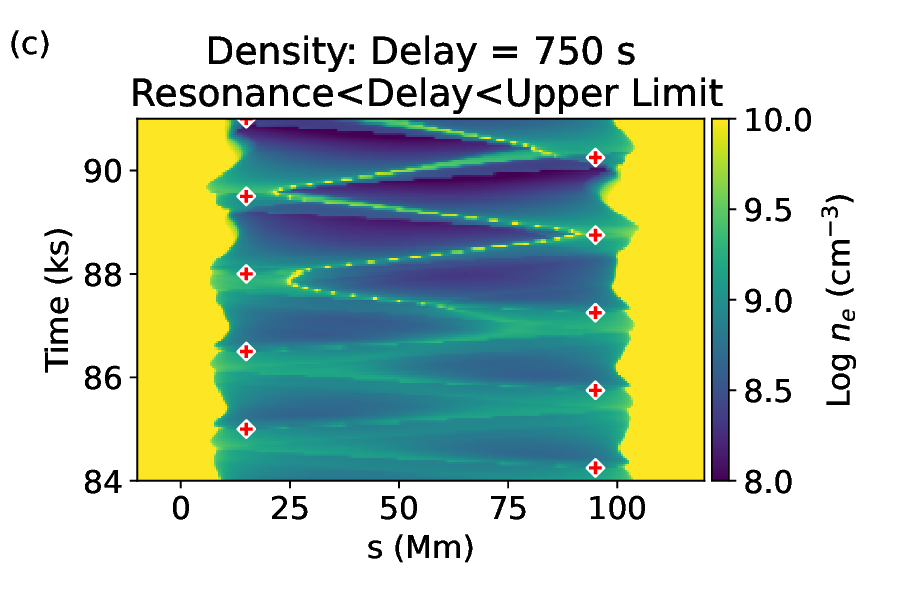}
  \includegraphics[height=5.5cm]{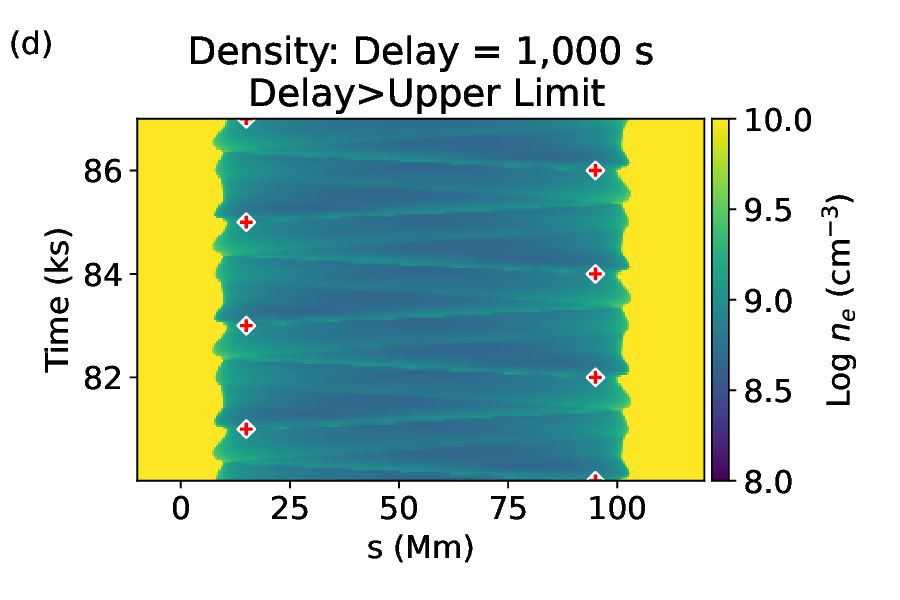}
\caption{Density as a function of location along the loop and time for four model runs with nanoflares alternating between the two sides of the loop with regular delay times and locations, but variations in the delay time between model runs. (a) A simulation with $\tau=150$~s, below the resonance value, (b) $\tau$ equal to the resonance, 450~s, (c) $\tau=750$~s, greater than the resonance delay time, but below the upper limit for condensation formation, and (d) $\tau=1,000$~s is above the upper limit for condensation formation. The times and locations of the nanoflares are shown by the red plus signs (+). The intermittent appearance of the condensations, most noticeable in panel (c), is an artifact due to the output grid size in the time dimension.}
\label{f:alternating_density}
 \end{figure}

\begin{figure}
\includegraphics[height=6cm]{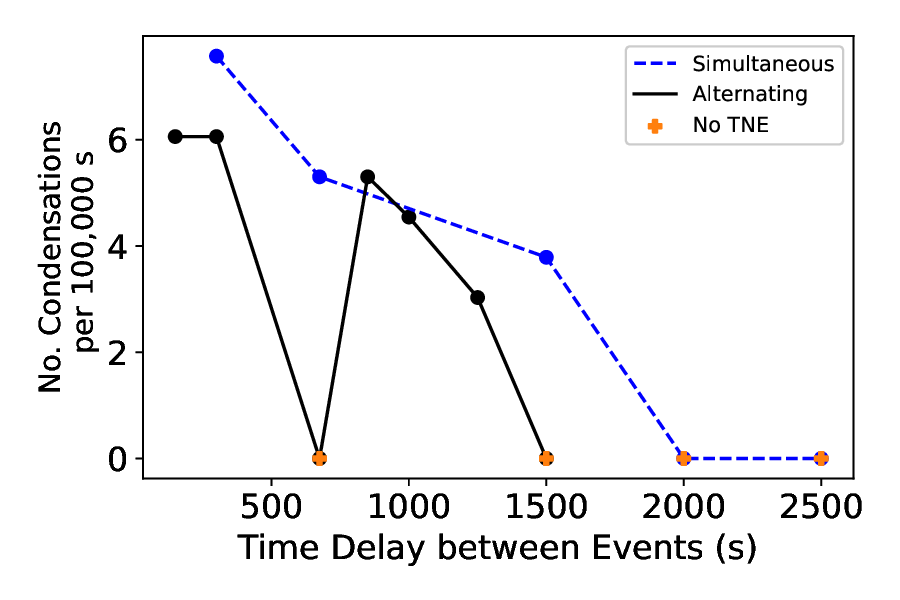}
\caption{Number of condensations per 100,000~s for runs with regularly occurring nanoflares occurring simultaneously in both legs (blue dashed line) or alternating between legs (black solid line) in a loop with $L_0=130$~Mm. Note the lack of condensations for the alternating nanoflare case at $\tau=675$~s. \revA{Data for this figure are in Table~\ref{t:NF7_sym_alt_ncond}.}}
\label{f:NF7_sym_alt_ncond}
\end{figure}

In order to investigate the idea that the  gap in condensation formation for the alternating nanoflares cases is due to this resonance we also simulated a longer loop, with $L_0=130$~Mm as opposed to the 80~Mm used for most of our runs (see Section~\ref{s:loop_geom}).
As shown in Figure~\ref{f:NF7_sym_alt_ncond} \revA{(and Table~\ref{t:NF7_sym_alt_ncond})}, we found that for the longer loop the gap in condensation formation also occurred, but at longer delay times, around $\tau=675$~sec, as opposed to 450~s for the shorter loop. Thus we see that the resonance delay time scales with loop length, as it should if the resonance conditions are determined by the evaporation front travel time along the loop.

 \subsubsection{Nanoflare Distributions with Randomized Cadence and Intensity}
 \label{s:rand_delay}
\begin{figure}
\includegraphics[height=6cm]{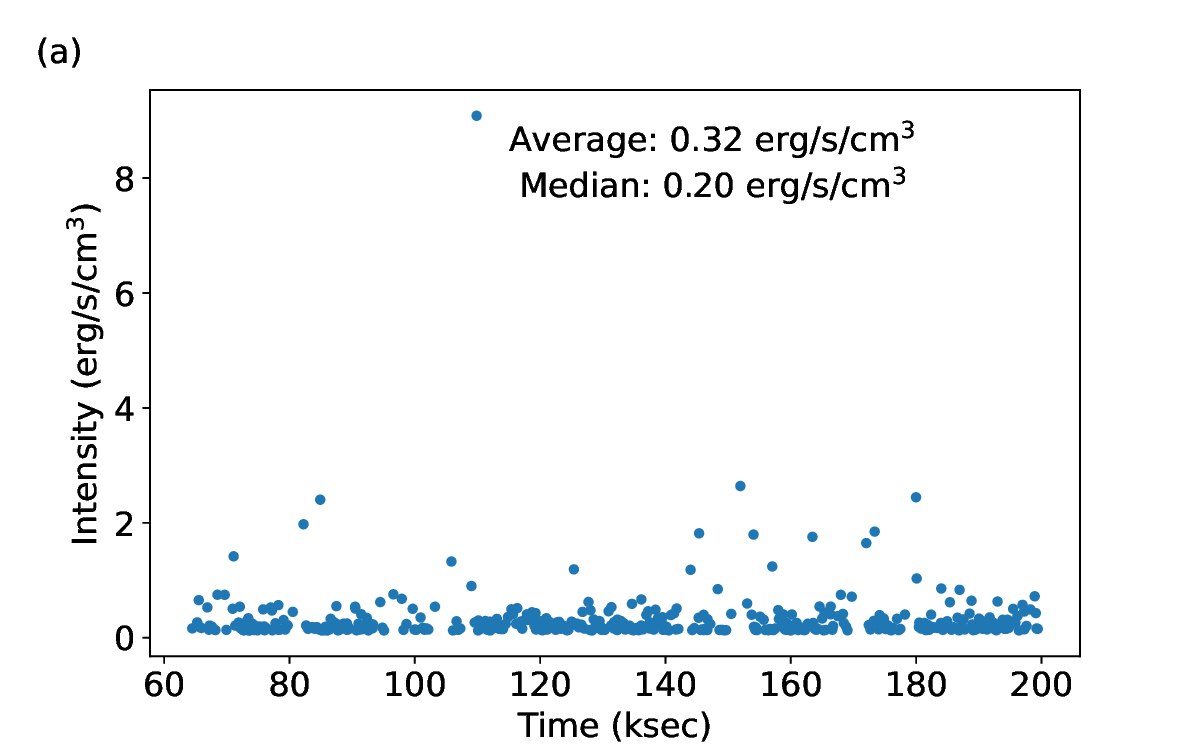}
\includegraphics[height=6cm]{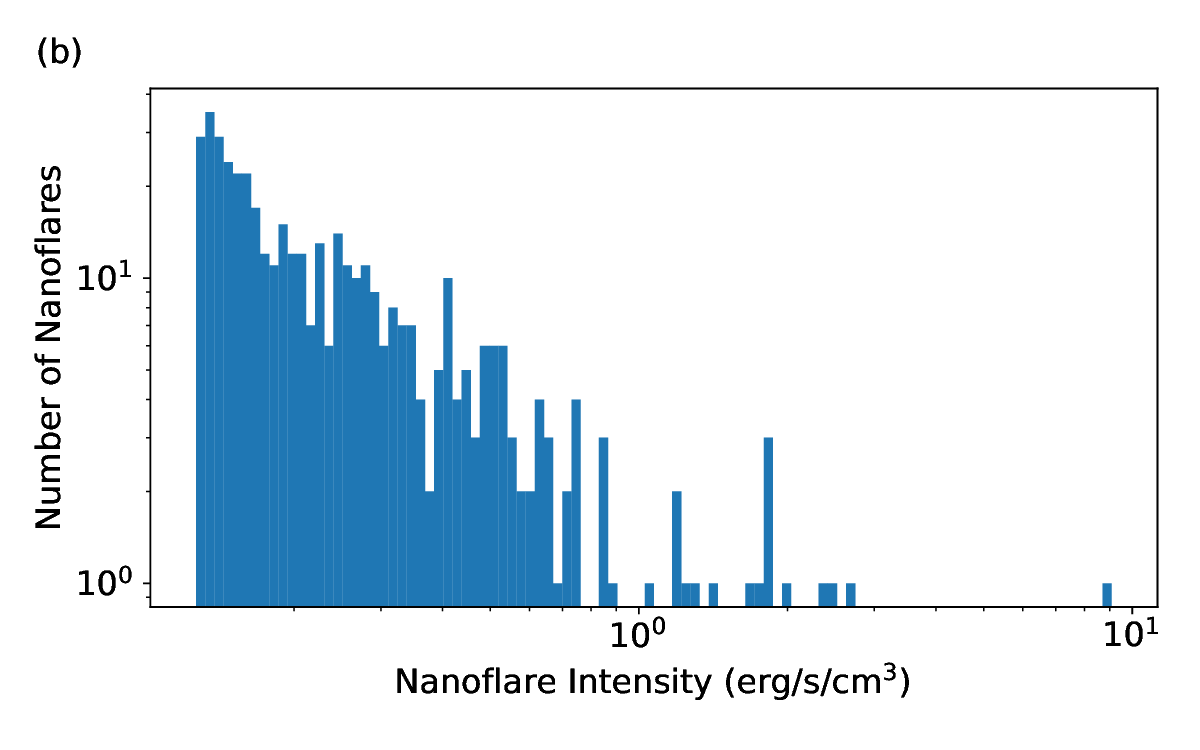}\\
\includegraphics[height=6cm]{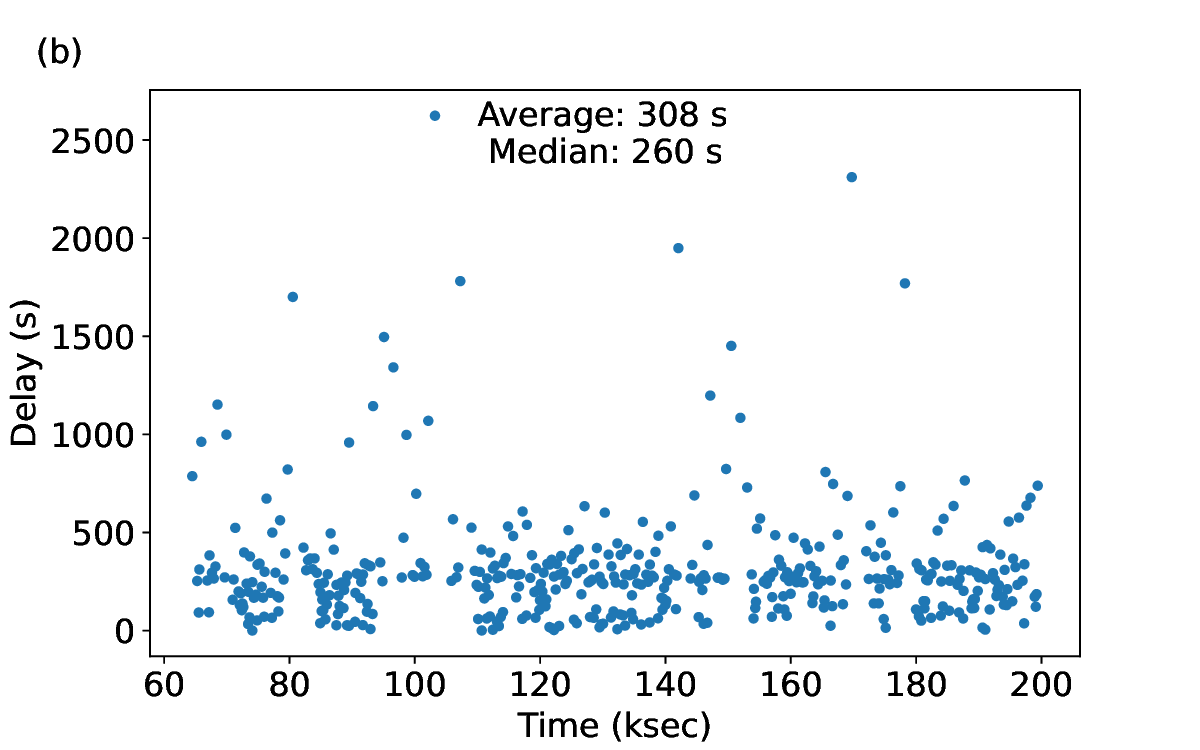}
\includegraphics[height=6cm]{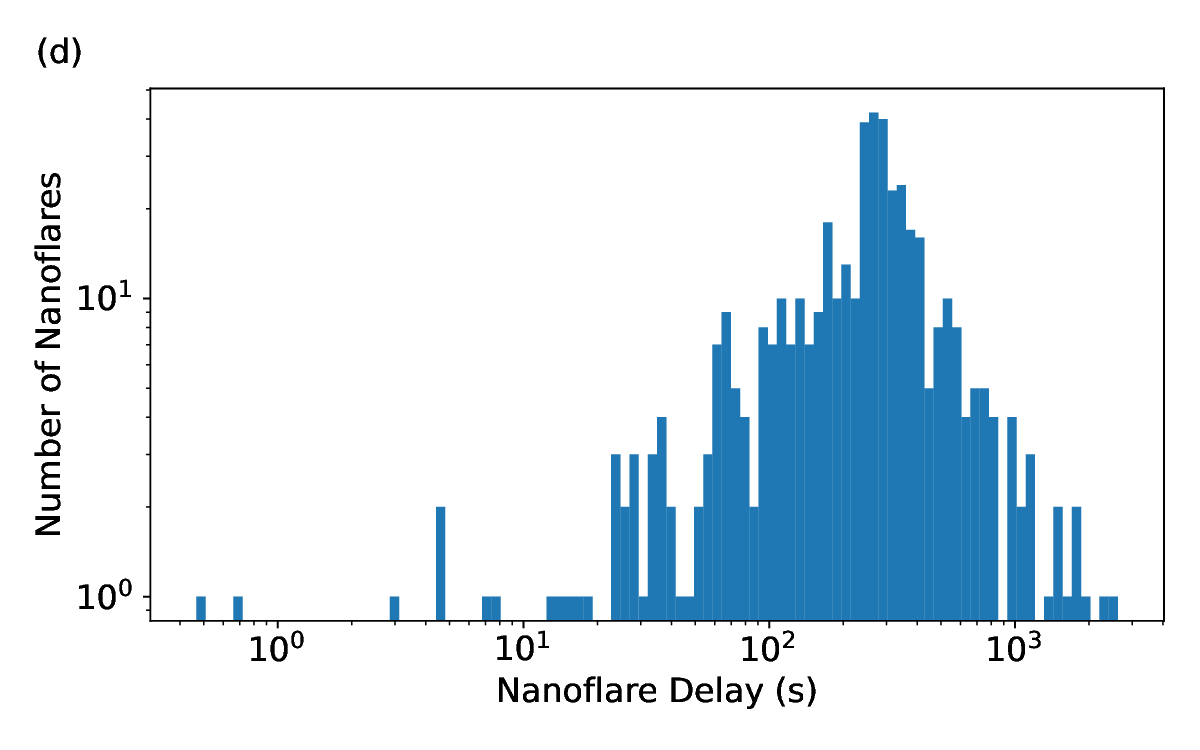}
\caption{Random power-law distribution of nanoflares with $\tavg = 308$~s, and average intensity of 0.32 erg~s$^{-1}$\cc\ and median of 0.2 erg~s$^{-1}$\cc \revA{over the entire loop. Independent distributions with similar properties are in both legs. (a) Intensity distribution as a function of time, (b) histogram of intensity distribution, (c) delay values as a function of time, (d) histogram of delay values.}}
\label{f:random_delay_dist}
\end{figure}

 More potentially realistic than cases with regular delay times are loops with random distributions of nanoflares. In this section we consider simulations with constant location, but randomized delays and related intensities, $\tau$ and $I\nf$. Values of $I\nf$ were first drawn randomly from a power-law probability distribution of slope -2.4 over the range 0.02 - 2.0 erg~s$^{-1}\cc$. Values of $\tau$ were then generated by specifying a characteristic delay and setting the specific delay between successive nanoflares to be proportional the magnitude of the first event. This corresponds to a physical picture of coronal magnetic strands being slowly tangled by photospheric convection and reconnecting when a critical misalignment angle between adjacent strands is reached, as modeled by \citet{lopez_15}, who found distributions of $\tau$ and $I\nf$ with the characteristics we use. Figure~\ref{f:random_delay_dist} shows an example, \revA{with the parameters both as a function of time and as histograms}. \revA{This is a combination of two power-law distributions with similar general properties, one in each leg of the loop.} The $I\nf$ values were normalized by setting the median $I\nf$ for the full randomized run equal to 0.2 erg~s$^{-1}\cc$ so that the $I\nf$ values would be similar to those of the loop-integrated nanoflare intensities in the simulations with constant intensity, \revA{although the average value will vary somewhat}. \revA{Because delay values are calculated based on the entire combined distribution from both legs, the histogram  (Fig.~\ref{f:random_delay_dist}d) is not a simple power-law.} As with the constant $\tau$ simulations discussed above, the nanoflares were located at $s=15$ and 95 Mm. For these simulations \davg\ ranged from 7.3 - 10.9~Mm, decreasing with increasing delay time.

\begin{figure}
\includegraphics[height=10cm]{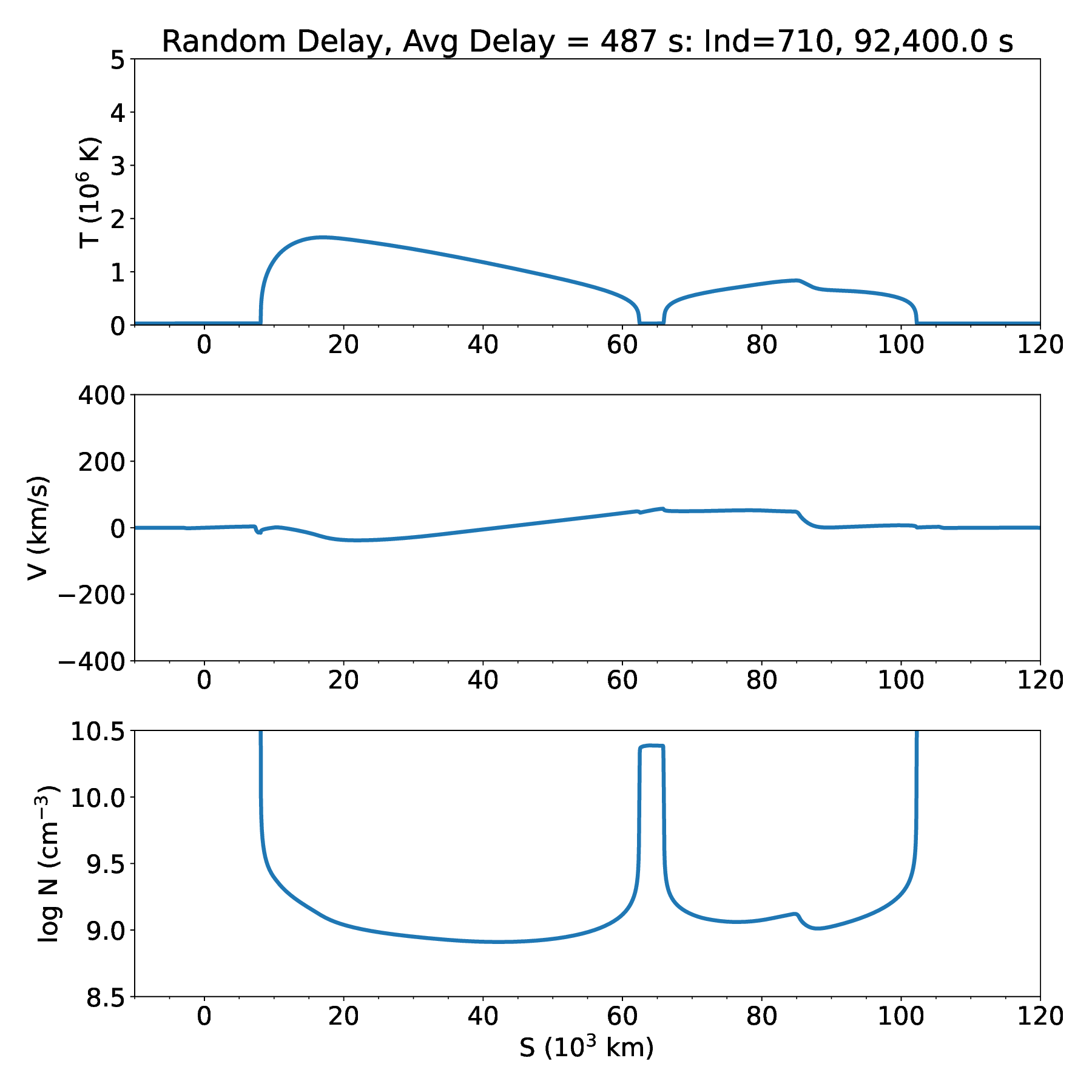} 
\caption{Temperature, velocity, and density along the loop for a simulation with nanoflares with randomized delays and intensities inversely proportional to the delay. $\tavg= 487$~s. A 19~s animation showing the simulation results from 74,000 - 97,000~s is included online.}
\label{f:cdip_NF4_21Jul2021_c2_frame}
\end{figure}

\begin{figure}
\includegraphics[height=5.5cm]{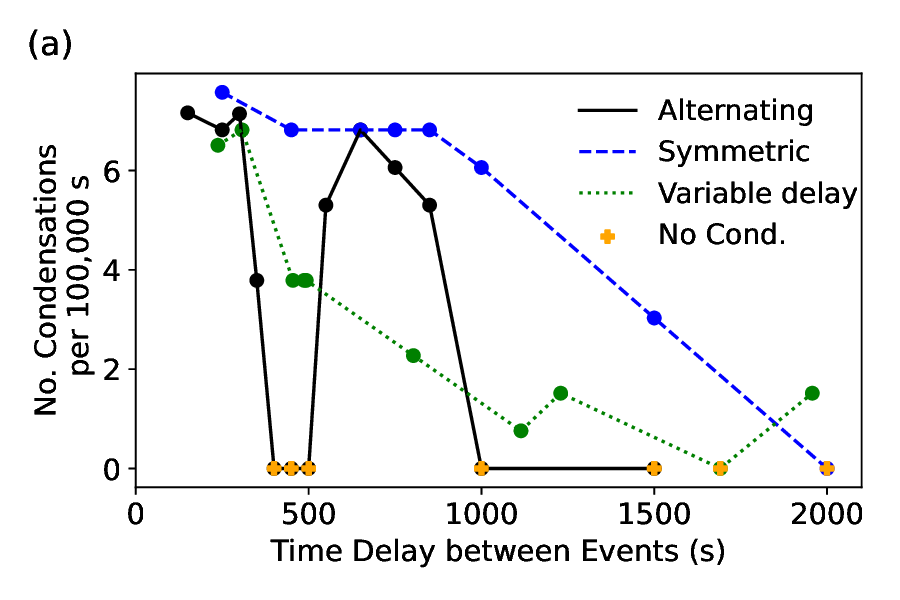}
\includegraphics[height=5.5cm]{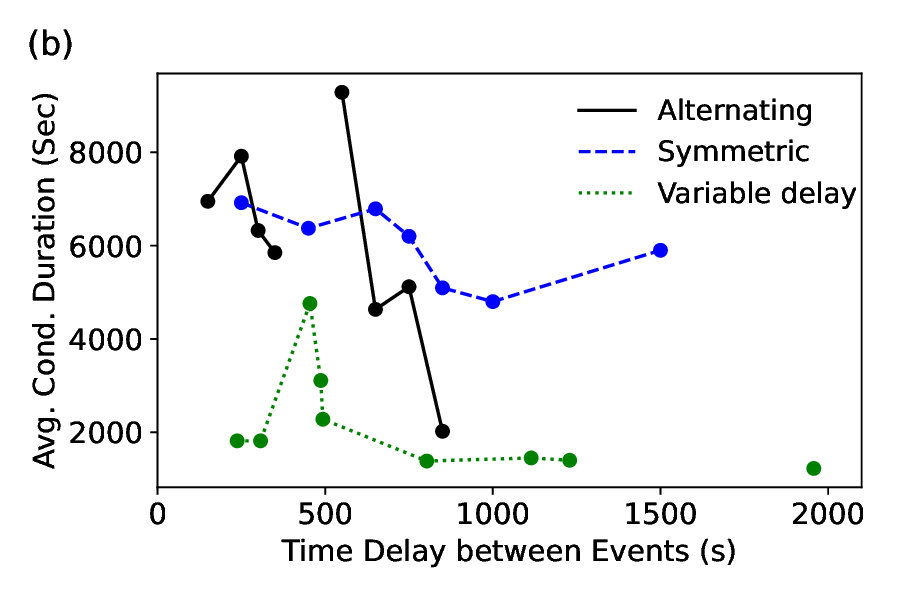}
\caption{Condensation characteristics for runs with regularly occurring nanoflares occurring simultaneously in both legs (blue dashed line) or alternating between legs (black solid line) and randomized delay times (green dotted) vs. average delay time. (a) The number of condensations per 100,000~s, (b) average duration of condensations over the run.
These runs were performed on a loop with $L_0=80$~Mm. \revA{Data for this figure are in Table~\ref{t:sym_alt_rand_ncond}.}}
\label{f:sym_alt_rand_ncond}
\end{figure}

The resulting condensations form over a wider range of locations within the loop than the condensations resulting from nanoflares with regular delays and can move back and forth along the loop a few times before exiting into the chromosphere. An example of this is shown in the animation accompanying Figure~\ref{f:cdip_NF4_21Jul2021_c2_frame}. As shown in Figure~\ref{f:sym_alt_rand_ncond}, there is not a gap in condensation formation near $\tavg=450$~s.  This is no surprise, since several regular cycles are necessary to set up a resonance. This indicates that it is necessary to have regular time intervals between nanoflares for this effect to occur. Condensations can occur with longer \tavg\ than seen for either the symmetric or alternating nanoflare cases. This extension of the critical  delay occurs because the random sequence always has the potential to have a run of nanoflares with $\tau$ shorter than what would be predicted based on \tavg.  \revA{The condensations occur at more irregular intervals reflecting the random nature of the nanoflare heating (see Figure~\ref{f:cond_count}b for an example.)}

However, in general, both the number of condensations and their duration are reduced compared to the simulations with constant $\tau$. Condensation requires multiple consecutive nanoflares that are closely enough spaced in time to produce an overall increase in density (and decrease in temperature) on top of the fluctuations associated with the individual events. The final thermal runaway that produces the condensation occurs when the cooling time finally drops below the draining time. With random nanoflares, there may not be enough closely spaced events to achieve these critical conditions before a long gap allows major draining to occur. The slow buildup must then restart from a low density. 

We also find that the durations of the condensations are much lower for the runs with randomized $\tau$ values. This is because the condensations are suspended in the corona by flows produced by subsequent nanoflares, and random distributions are likely to produce gaps between nanoflares during which the condensations to fall out of the loops.

 \subsection{Variations in Nanoflare Location}
\label{s:var_location}

Our procedure for varying the nanoflare location was similar to the variable delay cases. First we considered locations that were constant for each particular run but varied between runs, both with symmetric nanoflares in both legs and also for nanoflares alternating between legs of the loop. \revA{Then we performed runs in which we randomized the nanoflare locations.} For all runs  discussed in this sub-section $\tau=750$~s.
 
\subsubsection{Nanoflares at constant locations in the loop}
 \label{s:const_location}
 
 \begin{figure}
\includegraphics[height=5cm]{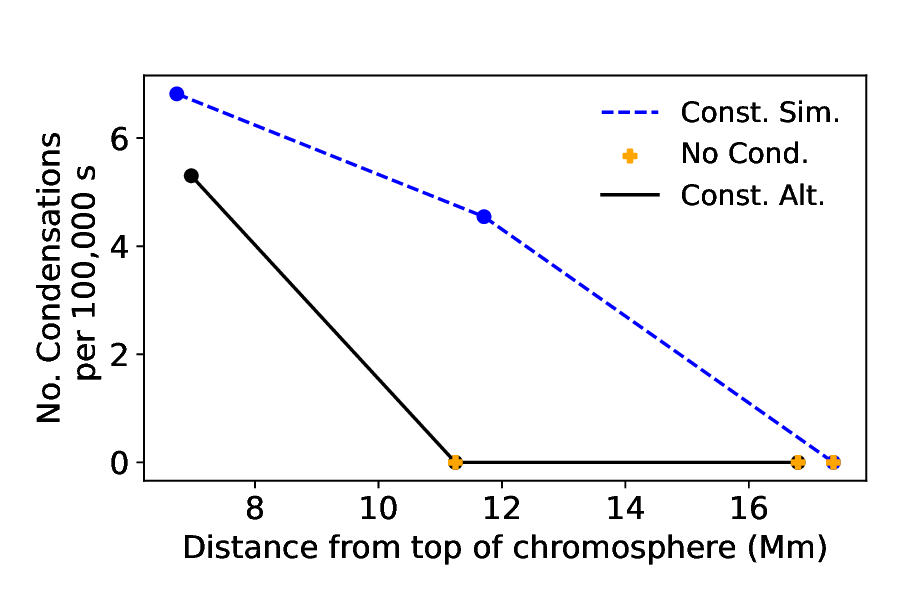}
\caption{The number of condensations per 100,000~s for runs with different fixed nanoflare locations. For all runs $\tau=750$~s. Shown are runs with (1) nanoflares occurring symmetrically and simultaneously in both legs (blue dashed line) and (2) alternating between legs at a constant distance from base of the feet (black solid line). The x-axis shows the average distance of the nanoflares from the top of the chromosphere at start of the nanoflare, excluding nanoflares centered at $< -1$ Mm below the top of the chromosphere. \revA{Data for this figure are in Table~\ref{t:sym_alt_rand_loc_ncond}.} }
\label{f:sym_alt_loc_ncond}
\end{figure} 

As can be seen from Figure~\ref{f:sym_alt_loc_ncond} \revA{(and Table~\ref{t:sym_alt_rand_loc_ncond})}, for the simulations with nanoflares occurring in the same locations over the entire run, the runs with heating closer to the top of the chromosphere produce more condensations, with a critical distance from the chromosphere, \dmax, above which nanoflares do not produce condensations \citep[e.g.,][]{testa_05,susino_10}. The cases with the simultaneous nanoflares in both legs produced more condensations than those with alternating nanoflares. The simultaneous nanoflares also have a higher \dmax. Both results, higher condensation rate and \dmax\ for simultaneous/symmetric nanoflares than for alternating ones, are consistent with our earlier discussion that colliding evaporative flows at the apex are more efficient at building up density.
 
 \subsubsection{Nanoflares at randomized locations in the loop}
  \label{s:rand_location}
\begin{figure}
\includegraphics[height=6cm]{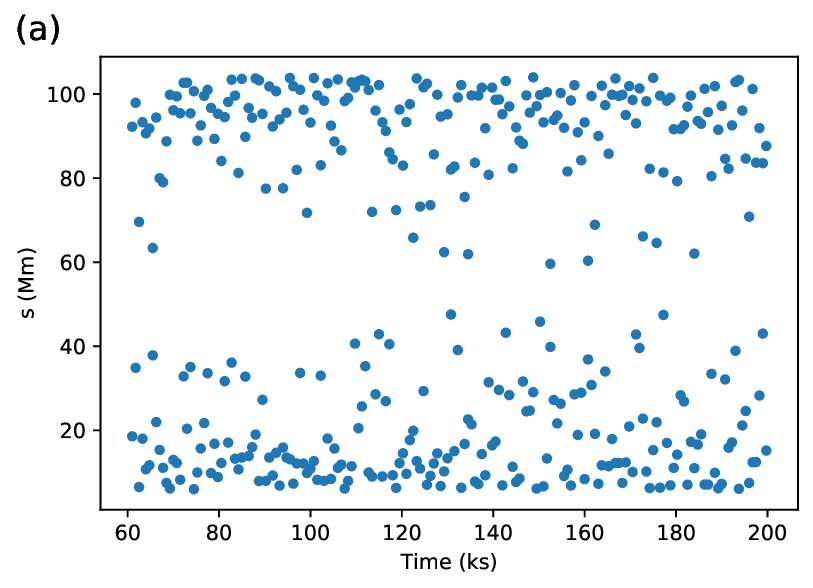}
\includegraphics[height=6cm]{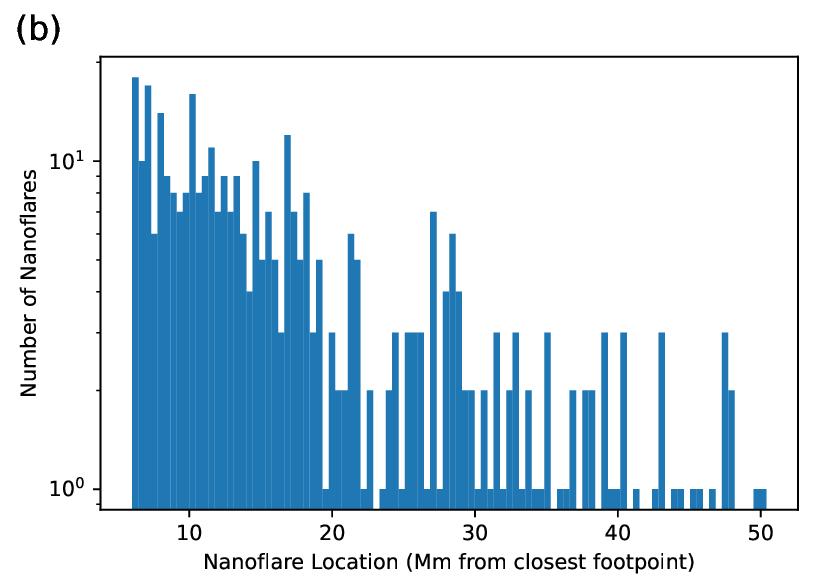}
\caption{Random exponential distribution of nanoflare locations in two legs starting at s=6~Mm with a scale height of 15~Mm \revA{as a function of time (panel (a)) and as a histogram as a function of the distance of the nanoflare from base of the closest foot (panel b). In this example the nanoflare time cadence is simultaneous in both legs and a constant 750~s.}}
\label{f:random_spatial_dist}
\end{figure}

Then we considered randomized locations in a given leg. Locations were selected randomly from a distribution that decreases exponentially with distance starting at s=6~Mm and having a specified scale length.  A distribution with similar characteristics, but not identical, was mirrored and applied to the other leg.  An example with a scale height of 15~Mm is shown in Figure~\ref{f:random_spatial_dist}.
We modeled cases in which nanoflares at these random locations were simultaneous, but not \revA{spatially} symmetric, in both legs, and cases in which they alternated.  Again, for all cases with simultaneous nanoflares in both legs the nanoflares had a peak intensity, $I\nf$, of 0.1~erg~s$^{-1}\cc$, while the alternating nanoflares had $I\nf=0.2$~erg~s$^{-1}\cc$ to maintain the same amount of input energy for each event.

As mentioned in Section~\ref{s:param} we are excluding nanoflares with peaks more than 1 Mm below the top of the chromosphere at the time of the nanoflare.  This is taken into account in our calculation for \davg, the average distance of the nanoflares from the top of the chromosphere. There is also some change in the effective time delay between events. In this section we are showing simulations with average input events delays of 750 s, but the effective delays are 750 - 762~s for the runs with simultaneous nanoflares in both legs and 784-857~s for the runs with nanoflares alternating between legs.

\begin{figure}
\includegraphics[height=5.5cm]{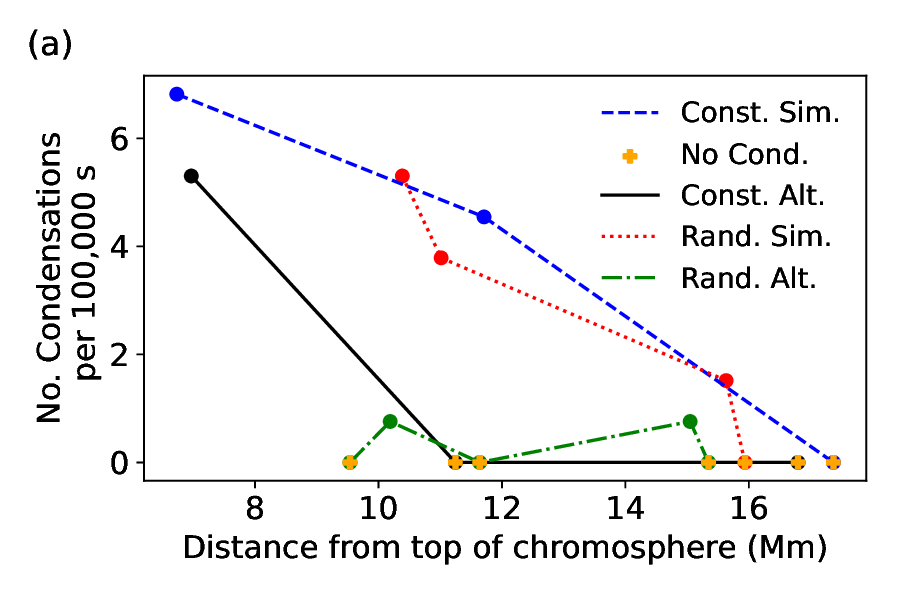}
\includegraphics[height=5.5cm]{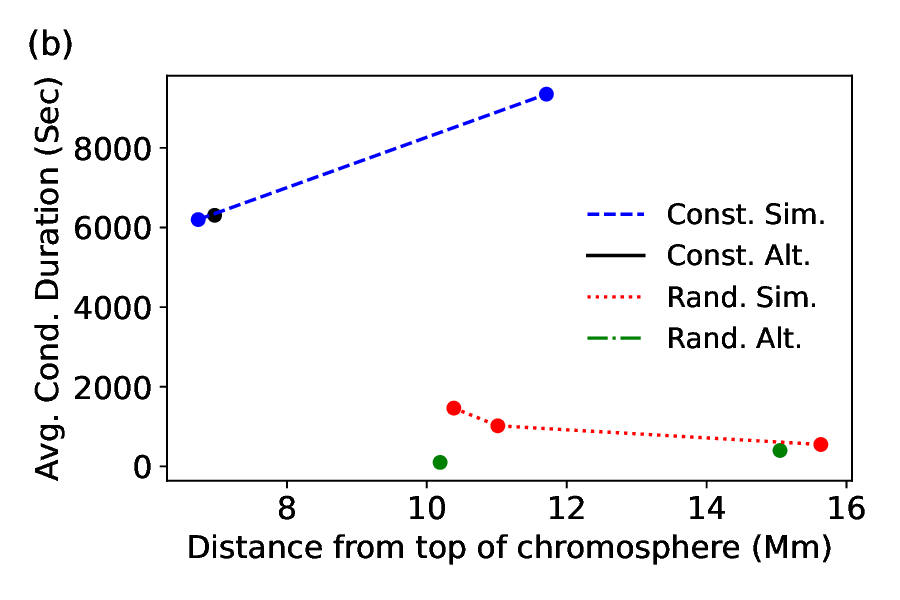}
\caption{Condensation characteristics for runs with varying nanoflare locations as a function of \davg. For all runs the delay between nanoflare events over the entire loop was 750~s. Shown are runs with nanoflares occurring symmetrically and simultaneously in both legs  (blue dashed line), alternating between legs at a constant distance from the base of the foot (black solid line), occurring simultaneously in both legs, but at random positions in each leg (red dotted), and alternating between legs but with randomized locations (green dash-dot). The plots show (a) the number of condensations per 100,000~s (b) average duration of condensations over the run. 
These runs were performed on a loop with $L_0=80$~Mm. \revA{Data for this figure are in Table~\ref{t:sym_alt_rand_loc_ncond}}.} 
\label{f:sym_alt_rand_loc_ncond}
\end{figure}

The number of condensations as a function \davg\ during the simulation run is shown in Figure~\ref{f:sym_alt_rand_loc_ncond} \revA{(and Table~\ref{t:sym_alt_rand_loc_ncond})}.  The randomized  runs with simultaneous nanoflares in both legs produce fewer condensations than the non-randomized symmetric runs, while the runs with alternating nanoflares and randomized locations usually produced the fewest condensations, although, as with delay randomization, the randomness means there are times in the sequence when the \davg\ exceeds the $\dmax$ for constant height nanoflares. As with randomization of $\tau$, randomization of nanoflare location reduces condensation formation because it reduces the efficiency of density build up, although the effect is smaller than when randomizing the delay.  The randomization also greatly reduces the time the condensations remain in the corona. As with the random time delays, random nanoflare locations are less likely to trap the condensations in the loop than ones occurring in a regularly occurring locations. 
 
 \subsection{Randomization of both delay time/intensity and location}
 \label{s:rand_delay_rand_loc}
 We finish by considering simulations with  both randomized delays/intensities and randomized locations. The nanoflare sequences used still had a roughly equal number of nanoflares in each leg. 
 
 \begin{figure}
\includegraphics[height=5cm]{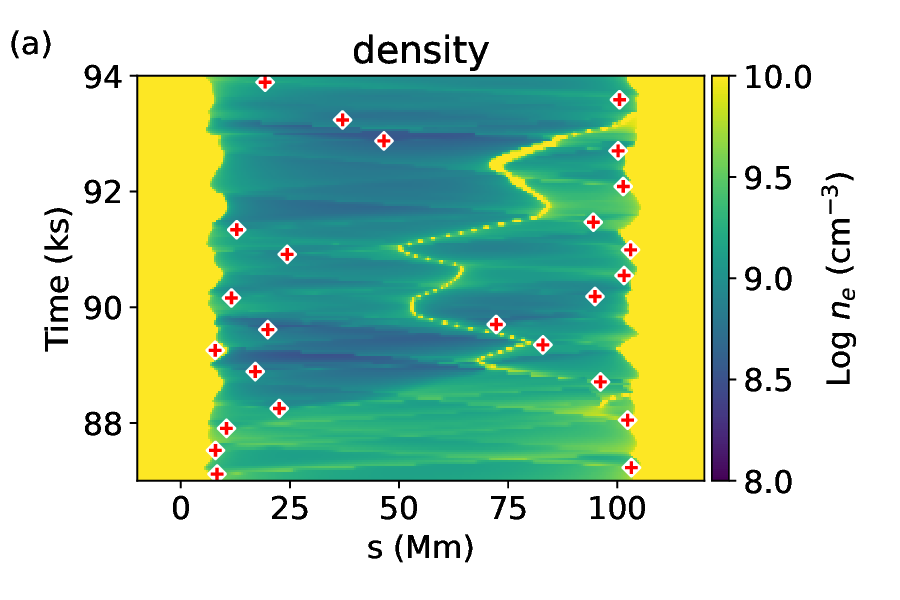}
\includegraphics[height=5cm]{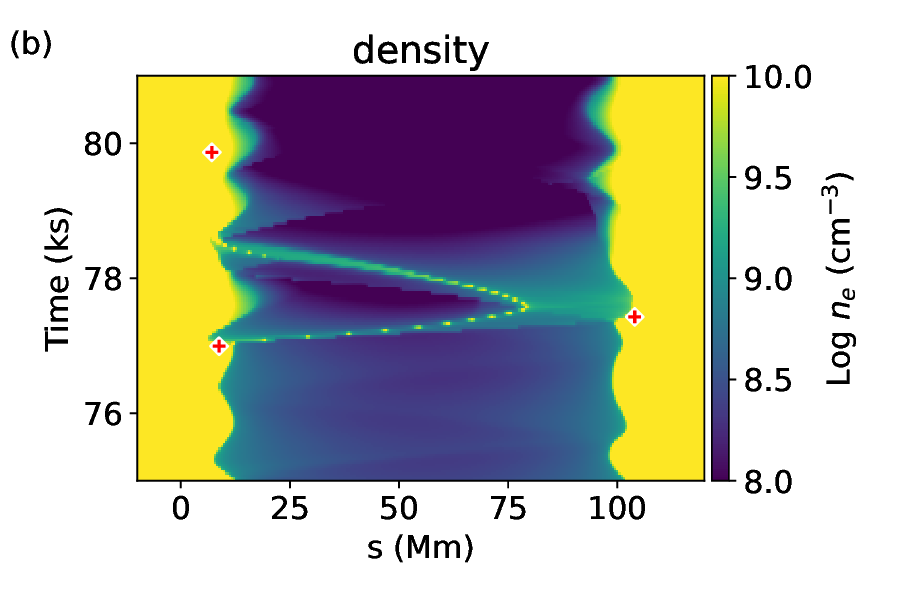}
\caption{Density as a function of time and distance along the loop for runs with random nanoflare time/intensity and random location. Red plus signs (+) show the times and locations of nanoflares. (a) A very short lived condensation followed by a long lived condensation forming in the corona from a simulation with $\tavg=469$~s. 
(b) A cool feature ejected from the chromosphere near the left foot and moving upwards from a simulation with $\tavg=2038$~s. Both simulations had $\davg =11.0$~Mm. }
\label{f:rand_rand_density}
\end{figure} 
 
 We found two general kinds of cool features in these simulations, shown in Figure~\ref{f:rand_rand_density}.  
  We saw condensations appearing in the loop, with life times depending on where in the loop they formed and how they were pushed back and forth by fronts produced by subsequent nanoflares.  An example is shown in Figure~\ref{f:rand_rand_density}a, where we see  the condensations  move back and forth in an erratic fashion in response to subsequent nanoflares.  Thus these sorts of complex, erratic  motions of condensations can be considered a sign of irregular impulsive heating in the loop.
 The condensations occur all over the loop - with ones forming closer to loop center being more common for the high cadence cases. \revA{The time between condensation formation is highly variable.}
 
\begin{figure}
\includegraphics[height=6cm]{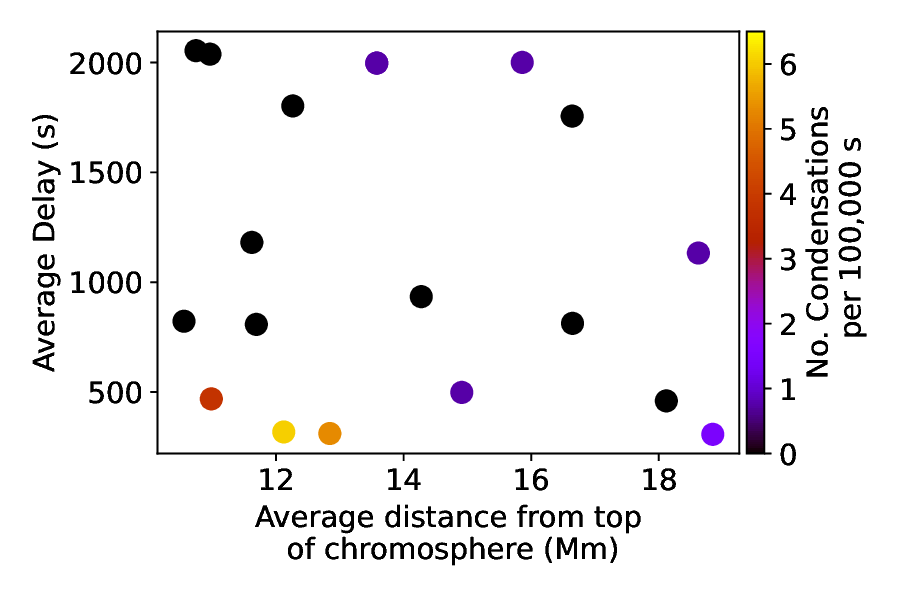}
\caption{The number of condensations as a function of average distance from the top of the chromosphere, \davg, and average delay, \tavg. \revA{Data for this figure are in Table~\ref{t:rand_delay_loc_ncond}}.}
\label{f:rand_delay_loc_ncond}
\end{figure}

 As we would expect, these condensations are more common for  simulations with low \tavg\ and \davg, as shown in Figure~\ref{f:rand_delay_loc_ncond} \revA{and Table~\ref{t:rand_delay_loc_ncond})}. If anything, the effect is stronger than in the cases in which only delay and intensity or only location were randomized.
Again, for cases with higher \tavg\ and \davg\ condensations become more rare, but occasionally occur due to short term, random decreases in $\tau$ and/or $d$. 

 In addition, we sometimes see cool features that appear to be ejected from the chromosphere in response to a nanoflare \revA{heating} immediately below the top the chromosphere  (Figure~\ref{f:rand_rand_density}b).  These move upwards and eventually also fall back down, usually on the same side of the loop. These seem to be distinct from the coronal condensations we are studying. They involve chromospheric material that preexists the nanoflare responsible for the ejection, not cool material that condenses out of the corona, and they occur in low numbers at higher values of \tavg\ and \davg. These show up in the random delay/random location runs because these simulations have a larger number of nanoflares submerged in the chromosphere than other runs. \revA{For purposes of automatic classification we classed a cool feature as an ejection rather than a condensation if it was first detected within 200~s and 10~Mm from a nanoflare that was centered no more than 2.25~Mm above the top of the chromosphere at the time it occurred}. Similar features have been observed in other modeling work, as we discuss in Section~\ref{s:discussion}. However, ARGOS, which incorporates the chromosphere in a relatively simple manner, is not the best tool to use to study such  ejections.

\section{Discussion}
\label{s:discussion}

Previous studies \citep[e.g.,][]{peres_93,testa_05} have utilized randomized nanoflares to model coronal loop heating  and found that these can still produce condensations, and generally follow the trends noted in our introduction (Section \ref{s:intro}) and also in our results.
\citet{mendoza-briceno_05}, for instance, state that for their simulations with randomized heating impulses, distributions of pulses closer to loop footpoints produce more rapid temperature depressions than do distributions extending further up the loop legs.  \citet{jercic_23} performed a study more focused on the parameters of the randomized nanoflare distribution, but applied to a two-dimensional prominence magnetic configuration rather than a one-dimensional loop configuration as we have here. They focused on variations in \revA{height} and intensity of the nanoflares.  
They found that their cases with ``high'' altitude nanoflares were more conducive to condensations, but the comparison was to nanoflares at lower altitudes that were often below the top of the chromosphere, so this is also consistent with our results in that we found that nanoflares below the top of the chromosphere had little effect on the loop (except for the ejecta discussed below), and thus essentially increased \tavg\ and discouraged condensations.  

\citet{antolin_08,antolin_10} and \citet{lix_22} \revA{utilized} randomized nanoflares with a focus on the dynamics of resulting condensations and  discuss erratic behavior in the condensations similar to what we see in our simulations. \citet{antolin_10} and \cite{sahin_23} report both upward as well as the more common downward motions in observations of coronal rain.  It may be the upward motions are associated with upward pressures from nanoflares that occur after the condensation forms, as in our models.

In our runs with random delay and random location (Section~\ref{s:rand_delay_rand_loc}) we noted that in addition to condensations we also found what seem to be ejecta from the chromosphere that sometimes occurred when the nanoflares were \revA{at least partially submerged in the chromosphere}.
A similar ejection phenomenon was observed in a simulation of randomized heating in a much smaller corona loop done by \citet{mendoza-briceno_06} who suggested they might represent spicules. \citet{huang_21} also discuss a dichotomy between the responses to heating events at different heights in the chromosphere near the footpoints of prominences in their simulations, suggesting that lower altitude energy input may be responsible for prominence formation via ejection, while energy injection close to the top of the chromosphere would be responsible for prominence formation via condensation. 
This is consistent with our results. However, the ARGOS model is not really appropriate for analyzing such features. They should be investigated more carefully using models incorporating MHD and a more sophisticated chromosphere, including radiative transfer effects.   

\revA{In general, simulations with randomized nanoflares exhibit a number of complex behaviors, including what appear to be aborted condensations that never fully fit our criteria before falling out of the loop and also periods when the loop appears to drain completely and then recovers to only the equilibrium level pre-nanoflare level. 
The first phenomenon might be compared to incomplete condensations of \citet{mikic_13} but seem different from the \citet{mikic_13} incomplete condensations which heat up again, whereas the ones in our simulation fall out of the loop. The highly dynamic conditions in our randomized nanoflare simulations make the comparison difficult. 
The second cooling and draining phenomenon is due to periods of long delays between nanoflares near the loop feet in the randomized sequences. }

Another limitation to our modeling is that it is confined to 1D processes. It is not able to incorporate cross loop phenomena like the MHD waves \citep{fang_15, xia_17b, claes_19} which are likely important in explaining long term, multi-loop phenomena such as the loop pulsations described by \citet{auchere_14, auchere_18,froment_15}. See the discussion in \citet{klimchuk_19}.

However, 1-D modeling as presented here can be used to develop further diagnostics of coronal heating. This was discussed by \citet{antolin_10} who point out, for instance,  that the \revA{spatially} uniform heating of loops expected from heating produced by Alfv\'en waves would tend to discourage condensations. On the other hand, the phenomenological wave-driven-turbulence (WDT) heating model of \citet{downs_16} has heating sufficiently concentrated at low altitudes to produce copious condensations. The heating is quasi-steady, whereas real turbulence heating is expected to be temporally intermittent, but the difference is unimportant if the frequency is sufficiently high. We know from the work presented here that the randomization of nanoflares, which we expect to be the most realistic situation, generally increases the already known restrictions that condensation implies that heating occurs with short cadence and near loop feet. It should also be possible to use simulations to investigate diagnostics such as thermal time-lags, which can vary in the presence condensations \citep[see][]{viall_20,froment_20} as well as differential emission measure.  Simulations can also be used to evaluate the likely effects of particular nanoflare distributions produced using physical principles \citep[e.g.,][]{knizhnik_20a,knizhnik_20b,knizhnik_22a}.

\section{Summary and Conclusions}
\label{s:summary}
In general, we find that condensations can form with nanoflare heating.
However, the occurrence rate is reduced in the more realistic cases when nanoflare timing and/or location are random, making the requirement that the delay times be short and the heating be near the feet even more important than for constant cadence and/or stationary heating sources.
That said, because with random sequences it is always possible to have a series of nanoflares with much shorter than average delay times or lower than average altitudes, it is always possible that a set of random nanoflares can occasionally produce a condensation assuming the simulation or observing period is long enough. \revA{Condensations appear with a regular cadence in simulations with constant nanoflare frequencies and locations. In contrast, randomization of nanoflare locations results in irregular condensation occurrence.}

 Trajectories of condensations in runs with randomized nanoflares are more complex than for nanoflares with regular distributions in frequency and location because the cool features can form \revA{at} any location of the loop and are pushed back and forth by the forces produced by subsequent randomly occurring nanoflares. Such motions are seen in coronal loops \citep{antolin_10, sahin_23}. Based on our results, such erratic behavior is a signature of impulsive heating.

Condensation lifetimes are also reduced by randomization as the bilateral symmetry (even for alternating nanoflares)  tends to keep  condensations in the loop;  this does not occur in randomized simulations in which evaporative flows produced by nanoflares may even push cool material out of the loop as opposed to keeping it in the corona.

We also find a that there exists a gap in condensation formation for the modeled case in which the cadence of regular alternating nanoflares corresponds with the time it takes the resulting flows to reach the other side of the loop. As with all highly regular nanoflare distributions, this is not likely to occur on the Sun, but is interesting from a perspective of understanding condensation formation.

Our future plans are to explore diagnostics like DEM distributions and thermal time lags, and to incorporate into the simulations the distributions of nanoflares that occur in MHD models driven by boundary motions meant to represent photospheric convection.

%

\begin{acknowledgments}

T.\ Kucera and J.\ Klimchuk were supported by the GSFC Heliophysics Internal Scientist Funding Model (competitive work package program), 
M.\ Luna acknowledges support through the Ram\'on y Cajal fellowship RYC2018-026129-I from the Spanish Ministry of Science and Innovation, the Spanish National Research Agency (Agencia Estatal de Investigaci\'on), the European Social Fund through Operational Program FSE 2014 of Employment, Education and Training and the Universitat de les Illes Balears. This publication is part of the R+D+i project PID2020-112791GB-I00, financed by MCIN/AEI/10.13039/501100011033. The authors thank E.\ Mason and P.\ MacNeice for help with the ARGOS code \revA{and C.\ Johnston for useful discussions. JAK benefited from participation on the International Space Science Institute team “Observe Local Think Global:  What Solar Observations can teach us about Mulitphase Plasmas across Astrophysical Scales” led by P.~Antolin and C.~Froment. We also thank the referee for suggestions that improved the paper.}
  
\end{acknowledgments}

\bibliographystyle{apj}
\bibliography{tak_bibliography}

\appendix
\section{Tables}
\label{s:tables}
\revA{In this appendix we present tables containing parameters pertaining to the different model runs discussed in this paper. Each one lists data values shown in one or more figures in the text. Listed are:}
\revA{
\begin{description}
\item[Delay/Intensity type]
 ``Constant'' for a single value over the course of the run or  ``Variable'' for a randomized distribution.
\item[$\tau_{avg}$] 
  Delay between events over the whole loop (not just a single leg). For Constant delay simulations this is a constant, for randomized distributions this is the average.  Nanoflare that were mostly below the top of the chromosphere were excluded (see Section~\ref{s:param}).
\item[Relative Timing]
 ``Simultaneous'' for simultaneous nanoflares in both legs, ``Alternating'' for alternating nanoflares between legs, and ``Variable'' for randomized delays between legs.
\item[$\Sigma\Imed$] Median of the sum of the peak intensities of nanoflares occurring simultaneously over the entire loop. If there is only one nanoflare at a time, as is the case for the Alternating and Variable timing runs, this is equal to the median peak energies of single nanoflares; otherwise, it the sum of the energies of the two simultaneous nanoflares.
\item[$\Sigma\Iavg$] Average of the sum of the peak intensities of nanoflares occurring simultaneously over the entire loop. Again, if there is only one nanoflare at a time, as is the case for the Alternating and Variable timing runs, this is equal to the average peak energy of single nanoflares; otherwise, it the sum of the energies of the two simultaneous nanoflares.
\item[Location Type] 
  ``Constant'' for a constant nanoflare location over the course of the model run, ``Variable'' for a randomized location distribution.
\item[\snf] 
  Distance of nanoflare centers from the top of the loop foot based on the initial model set-up. This is an average for randomized location distributions.
\item[$d_{avg}$] 
Average distance of nanoflare centers from the top of the chromosphere at the time they occurred. Nanoflare that were mostly below the top of the chromosphere were excluded (see Section~\ref{s:param}).
\item[Condensations per 100 ks ]
the number of condensations in a run normalized to a 100,000~s run length.
\item[Condensation Duration]
 average lifetime of condensations before they fall out of the loop.
\end{description}
}

\begin{table}
\caption{Simulation results for $L_0=80$~Mm, varying nanoflare delay times}
\begin{tabular}{llllllllll}
\toprule
Delay/ & \tavg & Relative & $\Sigma\Imed$ &$\Sigma\Iavg$ & Location & \snf & \davg & Cond. & Cond. \\
Intensity & (s) & Timing & (erg~s$^{-1}$ & (erg~s$^{-1}$ & Type & (Mm) & (Mm) & per & Duration \\
Type &  &  & \cc) & \cc) &  &  &  & 100 ks & (ks) \\
\hline
Constant & 250 & Simultaneous & 0.20 & 0.20 & Constant & 15.0 & 8.4 & 7.6 & 6.9 \\
Constant & 450 & Simultaneous & 0.20 & 0.20 & Constant & 15.0 & 7.9 & 6.8 & 6.4 \\
Constant & 650 & Simultaneous & 0.20 & 0.20 & Constant & 15.0 & 7.2 & 6.8 & 6.8 \\
Constant & 750 & Simultaneous & 0.20 & 0.20 & Constant & 15.0 & 6.7 & 6.8 & 6.2 \\
Constant & 850 & Simultaneous & 0.20 & 0.20 & Constant & 15.0 & 6.5 & 6.8 & 5.1 \\
Constant & 1000 & Simultaneous & 0.20 & 0.20 & Constant & 15.0 & 6.6 & 6.1 & 4.8 \\
Constant & 1500 & Simultaneous & 0.20 & 0.20 & Constant & 15.0 & 5.8 & 3.0 & 5.9 \\
Constant & 2000 & Simultaneous & 0.20 & 0.20 & Constant & 15.0 & 5.9 & 0.0 & N/A \\
Constant & 150 & Alternating & 0.20 & 0.20 & Constant & 15.0 & 9.2 & 7.2 & 7.0 \\
Constant & 250 & Alternating & 0.20 & 0.20 & Constant & 15.0 & 8.4 & 6.8 & 7.9 \\
Constant & 300 & Alternating & 0.20 & 0.20 & Constant & 15.0 & 8.0 & 7.1 & 6.3 \\
Constant & 350 & Alternating & 0.20 & 0.20 & Constant & 15.0 & 9.2 & 3.8 & 5.9 \\
Constant & 400 & Alternating & 0.20 & 0.20 & Constant & 15.0 & 12.5 & 0.0 & N/A \\
Constant & 450 & Alternating & 0.20 & 0.20 & Constant & 15.0 & 13.0 & 0.0 & N/A \\
Constant & 500 & Alternating & 0.20 & 0.20 & Constant & 15.0 & 16.2 & 0.0 & N/A \\
Constant & 550 & Alternating & 0.20 & 0.20 & Constant & 15.0 & 8.3 & 5.3 & 9.3 \\
Constant & 650 & Alternating & 0.20 & 0.20 & Constant & 15.0 & 7.5 & 6.8 & 4.6 \\
Constant & 750 & Alternating & 0.20 & 0.20 & Constant & 15.0 & 7.0 & 6.1 & 5.1 \\
Constant & 750 & Alternating & 0.20 & 0.20 & Constant & 15.0 & 7.0 & 5.3 & 6.3 \\
Constant & 850 & Alternating & 0.20 & 0.20 & Constant & 15.0 & 6.5 & 5.3 & 2.0 \\
Constant & 1000 & Alternating & 0.20 & 0.20 & Constant & 15.0 & 5.8 & 0.0 & N/A \\
Constant & 1500 & Alternating & 0.20 & 0.20 & Constant & 15.0 & 10.8 & 0.0 & N/A \\
Variable & 238 & Variable & 0.20 & 0.38 & Constant & 15.0 & 10.9 & 6.5 & 1.8 \\
Variable & 307 & Variable & 0.20 & 0.32 & Constant & 15.0 & 9.7 & 6.8 & 1.8 \\
Variable & 445 & Variable & 0.20 & 0.34 & Constant & 15.0 & 9.0 & 1.8 & 3.3 \\
Variable & 455 & Variable & 0.20 & 0.35 & Constant & 15.0 & 8.9 & 3.8 & 4.8 \\
Variable & 487 & Variable & 0.20 & 0.36 & Constant & 15.0 & 8.9 & 3.8 & 3.1 \\
Variable & 493 & Variable & 0.20 & 0.32 & Constant & 15.0 & 8.7 & 3.8 & 2.3 \\
Variable & 803 & Variable & 0.21 & 0.31 & Constant & 15.0 & 8.1 & 2.3 & 1.4 \\
Variable & 1114 & Variable & 0.21 & 0.36 & Constant & 15.0 & 8.1 & 0.8 & 1.5 \\
Variable & 1229 & Variable & 0.19 & 0.32 & Constant & 15.0 & 8.1 & 1.5 & 1.4 \\
Variable & 1691 & Variable & 0.21 & 0.39 & Constant & 15.0 & 7.3 & 0.0 & N/A \\
Variable & 1957 & Variable & 0.18 & 0.35 & Constant & 15.0 & 7.3 & 1.5 & 1.2 \\
\hline
\end{tabular}
\tablecomments{Includes data for Figures \ref{f:sym_alt_ncond} and \ref{f:sym_alt_rand_ncond}.}
\label{t:sym_alt_rand_ncond}
\end{table}

\begin{table}
\caption{Simulation results for $L_0=135$~Mm, varying nanoflare delay times}
\begin{tabular}{llllllllll}
\toprule
Delay/ & \tavg & Relative & $\Sigma\Imed$ &$\Sigma\Iavg$  & Location & \snf & \davg & Cond. & Cond. \\
Intensity & (s) & Timing & (erg~s$^{-1}$ & (erg~s$^{-1}$ & Type & (Mm) & (Mm) & per & Duration \\
Type &  &  & \cc) & \cc) &  &  &  & 100 ks & (ks) \\
\hline
Constant & 300 & Simultaneous & 0.20 & 0.20 & Constant & 15.0 & 8.3 & 7.6 & 5.6 \\
Constant & 675 & Simultaneous & 0.20 & 0.20 & Constant & 15.0 & 8.4 & 5.3 & 6.7 \\
Constant & 1500 & Simultaneous & 0.20 & 0.20 & Constant & 15.0 & 6.5 & 3.8 & 6.0 \\
Constant & 2000 & Simultaneous & 0.20 & 0.20 & Constant & 15.0 & 7.4 & 0.0 & N/A \\
Constant & 2500 & Simultaneous & 0.20 & 0.20 & Constant & 15.0 & 6.9 & 0.0 & N/A \\
Constant & 150 & Alternating & 0.20 & 0.20 & Constant & 15.0 & 9.5 & 6.1 & 7.2 \\
Constant & 300 & Alternating & 0.20 & 0.20 & Constant & 15.0 & 9.0 & 6.1 & 6.8 \\
Constant & 675 & Alternating & 0.20 & 0.20 & Constant & 15.0 & 11.1 & 0.0 & N/A \\
Constant & 850 & Alternating & 0.20 & 0.20 & Constant & 15.0 & 7.9 & 5.3 & 5.2 \\
Constant & 1000 & Alternating & 0.20 & 0.20 & Constant & 15.0 & 7.5 & 4.5 & 4.5 \\
Constant & 1250 & Alternating & 0.20 & 0.20 & Constant & 15.0 & 6.9 & 3.0 & 4.4 \\
Constant & 1500 & Alternating & 0.20 & 0.20 & Constant & 15.0 & 6.8 & 0.0 & N/A \\
\hline
\end{tabular}
\tablecomments{Includes data for Figure \ref{f:NF7_sym_alt_ncond}.  }
\label{t:NF7_sym_alt_ncond}
\end{table}

\begin{table}
\caption{Simulation results for $L_0=80$~Mm varying nanoflare locations}
\begin{tabular}{llllllllll}
\toprule
Delay/ & \tavg & Relative & $\Sigma\Imed$ &$\Sigma\Iavg$  & Location & \snf & \davg & Cond. & Cond. \\
Intensity & (s) & Timing & (erg~s$^{-1}$ & (erg~s$^{-1}$ & Type & (Mm) & (Mm) & per & Duration \\
Type &  &  & \cc) & \cc) &  &  &  & 100 ks & (ks) \\
\hline
Constant & 750 & Simultaneous & 0.20 & 0.20 & Constant & 15.0 & 6.7 & 6.8 & 6.2 \\
Constant & 750 & Simultaneous & 0.20 & 0.20 & Constant & 20.0 & 11.7 & 4.5 & 9.3 \\
Constant & 750 & Simultaneous & 0.20 & 0.20 & Constant & 25.0 & 17.4 & 0.0 & N/A \\
Constant & 750 & Alternating & 0.20 & 0.20 & Constant & 15.0 & 7.0 & 5.3 & 6.3 \\
Constant & 750 & Alternating & 0.20 & 0.20 & Constant & 20.0 & 11.2 & 0.0 & N/A \\
Constant & 750 & Alternating & 0.20 & 0.20 & Constant & 25.0 & 16.8 & 0.0 & N/A \\
Constant & 762 & Simultaneous & 0.20 & 0.20 & Variable & 17.0 & 10.4 & 5.3 & 1.5 \\
Constant & 758 & Simultaneous & 0.20 & 0.20 & Variable & 17.9 & 11.0 & 3.8 & 1.0 \\
Constant & 750 & Simultaneous & 0.20 & 0.20 & Variable & 22.7 & 15.6 & 1.5 & 0.6 \\
Constant & 750 & Simultaneous & 0.20 & 0.20 & Variable & 23.3 & 15.9 & 0.0 & N/A \\
Constant & 831 & Alternating & 0.20 & 0.20 & Variable & 16.8 & 9.5 & 0.0 & N/A \\
Constant & 857 & Alternating & 0.20 & 0.20 & Variable & 17.1 & 10.2 & 0.8 & 0.1 \\
Constant & 836 & Alternating & 0.20 & 0.20 & Variable & 18.6 & 11.6 & 0.0 & N/A \\
Constant & 784 & Alternating & 0.20 & 0.20 & Variable & 22.3 & 15.0 & 0.8 & 0.4 \\
Constant & 793 & Alternating & 0.20 & 0.20 & Variable & 22.6 & 15.3 & 0.0 & N/A \\
\hline
\end{tabular}
\tablecomments{Includes data for Figures \ref{f:sym_alt_loc_ncond} and \ref{f:sym_alt_rand_loc_ncond}.}
\label{t:sym_alt_rand_loc_ncond}
\end{table}

\begin{table}
\caption{Simulation results for $L_0=80$~Mm randomized distributions of nanoflare delays and locations}
\begin{tabular}{llllllllll}
\toprule
Delay/ & \tavg & Relative & $\Sigma\Imed$ &$\Sigma\Iavg$  & Location & \snf & \davg & Cond. & Cond. \\
Intensity & (s) & Timing & (erg~s$^{-1}$ & (erg~s$^{-1}$ & Type & (Mm) & (Mm) & per & Duration \\
Type &  &  & \cc) & \cc) &  &  &  & 100 ks & (ks) \\
\hline
Variable & 307 & Variable & 0.20 & 0.32 & Variable & 23.4 & 18.8 & 1.5 & 1.7 \\
Variable & 311 & Variable & 0.20 & 0.32 & Variable & 17.8 & 12.8 & 5.3 & 1.8 \\
Variable & 318 & Variable & 0.20 & 0.32 & Variable & 17.0 & 12.1 & 6.1 & 1.6 \\
Variable & 469 & Variable & 0.20 & 0.35 & Variable & 16.5 & 11.0 & 3.8 & 1.8 \\
Variable & 459 & Variable & 0.20 & 0.35 & Variable & 23.5 & 18.1 & 0.0 & N/A \\
Variable & 498 & Variable & 0.20 & 0.31 & Variable & 20.6 & 14.9 & 0.8 & 0.4 \\
Variable & 822 & Variable & 0.21 & 0.31 & Variable & 17.2 & 10.6 & 0.0 & N/A \\
Variable & 808 & Variable & 0.21 & 0.31 & Variable & 18.3 & 11.7 & 0.0 & N/A \\
Variable & 813 & Variable & 0.21 & 0.31 & Variable & 22.9 & 16.6 & 0.0 & N/A \\
Variable & 934 & Variable & 0.21 & 0.30 & Variable & 20.8 & 14.3 & 0.0 & N/A \\
Variable & 1181 & Variable & 0.21 & 0.36 & Variable & 17.8 & 11.6 & 0.0 & N/A \\
Variable & 1133 & Variable & 0.21 & 0.36 & Variable & 25.4 & 18.6 & 0.8 & 0.1 \\
Variable & 1802 & Variable & 0.21 & 0.39 & Variable & 19.1 & 12.3 & 0.0 & N/A \\
Variable & 1756 & Variable & 0.21 & 0.39 & Variable & 23.5 & 16.6 & 0.0 & N/A \\
Variable & 2054 & Variable & 0.20 & 0.32 & Variable & 17.0 & 10.7 & 0.0 & N/A \\
Variable & 2038 & Variable & 0.18 & 0.25 & Variable & 18.3 & 11.0 & 0.0 & N/A \\
Variable & 2001 & Variable & 0.19 & 0.29 & Variable & 22.9 & 15.9 & 0.8 & 0.2 \\
Variable & 1998 & Variable & 0.19 & 0.36 & Variable & 20.7 & 13.6 & 0.8 & 0.0 \\
\hline
\end{tabular}
\tablecomments{Includes data for Figure \ref{f:rand_delay_loc_ncond}.}
\label{t:rand_delay_loc_ncond}
\end{table}

\end{document}